\documentclass[preprint,11pt]{elsarticle}

\usepackage{amssymb}
\usepackage{amsmath}
\usepackage{array}
\usepackage{booktabs}

\renewcommand{\vec}[1]{\boldsymbol{#1}}

\usepackage{graphicx}
\usepackage{subcaption}
\usepackage{float}

\makeatletter
\def\ps@pprintTitle{}
\makeatother

\title{A Variational Formulation for Deformable Particle Simulations and its Level Set Discrete Element Method Implementation}

\author[EPFL]{Thomas Henzel}

\author[EPFL]{Konstantinos Karapiperis\corref{cor1}}

\affiliation[EPFL]{organization={Data-Driven Mechanics Laboratory, School of Architecture, Civil and Environmental Engineering, EPFL},
city={Lausanne},
country={Switzerland}}

\cortext[cor1]{Corresponding author.}

\begin{document}

\begin{frontmatter}

\begin{abstract}
We present a deformable Discrete Element Method (DEM) that extends the classical rigid-particle formulation through a reduced-order description of elastic grain-scale deformation. The method hinges on two developments. First, an energetic variational formulation based on the Lagrange--d'Alembert principle extends classical rigid-body dynamics to incorporate particle deformability by embedding translational, rotational, and deformation degrees of freedom within a unified energetic description. Second, particle deformation is realized within the Level Set DEM formalism through evolving level sets. The framework applies broadly to general particle geometries and topologies, and supports arbitrary deformation modes. The resulting deformable DEM retains the robustness, geometric and physical clarity, and scalability of classical DEM, while enabling physically grounded grain-scale deformability at a computational cost of the same order of magnitude as rigid DEM. Comparisons with full finite-element simulations demonstrate excellent agreement at both particle and system scales, establishing a general and extensible variational framework for modeling deformation in particulate systems.
\end{abstract}


\end{frontmatter}

\section{Introduction}
\label{Introduction}

Discrete particulate systems constitute a fundamental form of matter in science, nature, and engineering. \emph{Particulate matter} may be defined as material systems composed of discrete entities, termed particles, whose collective behavior arises from their particulate structure and inter-particle interactions, and cannot generally be reduced to a single continuous field description without loss of essential physics. Assemblies of interacting building blocks—whether composed of cohesively bonded or frictionally interacting units—arise across a wide range of physical contexts and length scales. Classical examples include geomaterials such as soils and rocks~\cite{donze2009advances}, as well as load-bearing and protective tissues of biological systems, such as nacre and fish scales~\cite{barthelat2011toughness}. Architected structural materials composed of bonded or frictionally interlocking discrete units constitute an important realization of particulate structures~\cite{feldfogel2024failure}. Particulate forms are prevalent in active and living matter as well, including biofilms~\cite{mattei2018continuum} and cancer tumors~\cite{anderson1998continuous}. Furthermore, industrial systems and processes rely heavily on granular forms of matter, such as pharmaceutical powders~\cite{yeom2019application}. Across all these systems, the emergent mechanical response is governed fundamentally by two inseparable ingredients: the interactions between constituent particles and the deformation and failure of the particles themselves. Thus, a computational framework for predictive modeling of such systems must explicitly capture both \textit{particle–particle interactions} and \textit{particle-level deformation}. At the same time, the framework must remain computationally efficient and scalable, enabling simulations of the very large assemblies characteristic of natural and engineered particulate matter.
\begin{figure}[t]
    \centering
    \includegraphics[width=\textwidth]{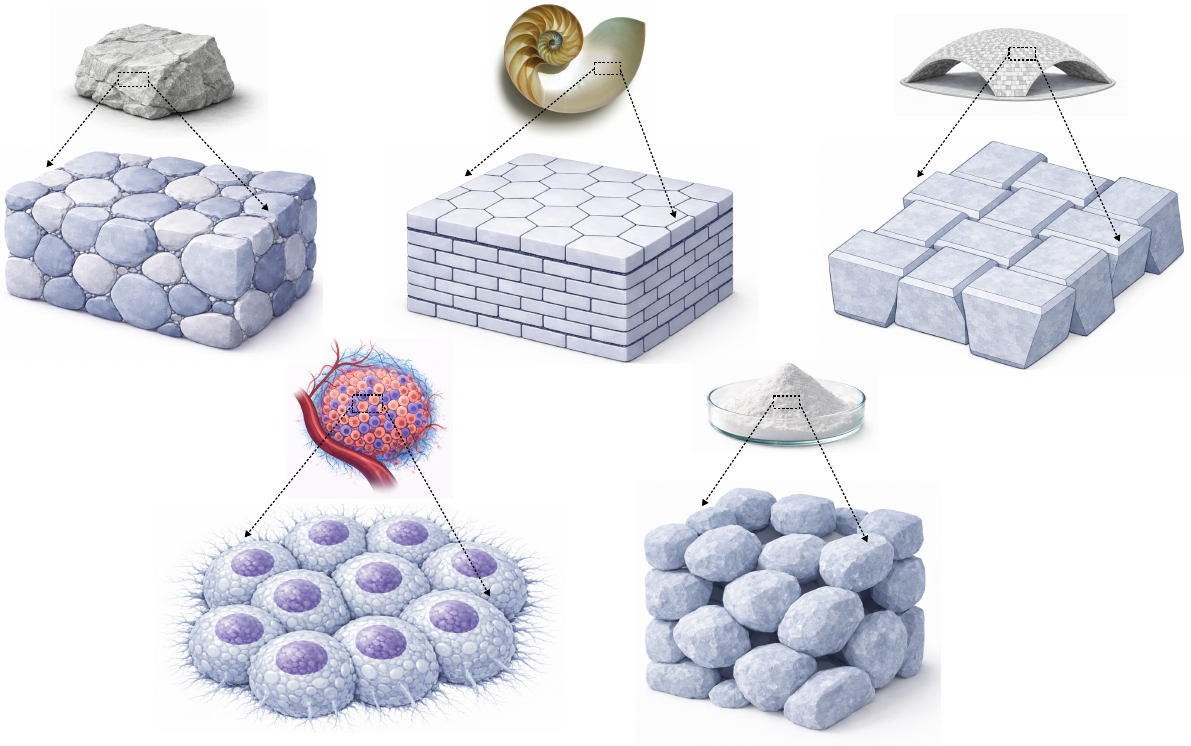}

\caption{Representative particulate systems across length scales and physical contexts, illustrating the breadth of \emph{particulate matter} considered in this work: material systems composed of discrete entities whose collective mechanical behavior arises from inter-particle interactions and particle deformation. Examples include geomaterials (e.g., rocks), biological lamellar composites (e.g., nacre), architected assemblies of bonded or frictionally interlocking discrete units (e.g., topologically interlocked structures), cellular aggregates in living matter (e.g., cancer tumors), and industrial granular systems (e.g., pharmaceutical powders). The figure highlights the unifying discrete mechanics perspective motivating the deformable DEM framework developed herein.}
    \label{fig:ParticulateMatter}
\end{figure}

The Finite Element Method (FEM) provides a high-fidelity framework for resolving particle-level deformations, but it struggles to model large assemblies of discrete blocks whose behavior is governed by nonlinear interactions such as contact and frictional sliding~\cite{munjiza2004combined}. These interactions introduce discontinuities and strong nonlinearities that severely limit the scalability of FEM, rendering it impractical for systems containing large numbers of particles~\cite{cundall1992numerical}. The Discrete Element Method (DEM), by contrast, is naturally formulated for such problems. In DEM, materials are represented as collections of discrete particles interacting through contact or bonded interfaces, making it exceptionally well suited for simulating large assemblies governed by discrete mechanical interactions~\cite{cundall1979discrete}. Traditional DEM relies on \emph{rigid} particles whose shapes remain fixed throughout the simulation. Although the contact laws used in DEM---through normal overlap, tangential slip, and appropriate stiffness parameters---can successfully approximate \emph{local deformation}, such as that described by Hertzian theory, these mechanisms do not produce \emph{bulk deformation}, that is, global shape changes of individual grains. In what follows, we use \emph{local deformation} to refer to the contact-level indentation or tangential slip implicitly captured by rigid DEM, and \emph{bulk deformation} to denote the overall elastic or inelastic shape change of a particle. Consequently, in order to faithfully model a wider range of particulate systems as discussed above, classical DEM—and, more generally, rigid-particle, contact-based formulations—must be extended to incorporate bulk deformation of the grains.

A number of approaches have been proposed in the literature to endow DEM particles with deformability, aiming to overcome the limitations of the classical rigid-particle formulation. One strategy is to treat each particle as a continuum body and resolve its deformation by solving an internal boundary value problem, typically via FEM~\cite{ghaboussi1988fully,barbosa1990discrete, munjiza1995combined,komodromos2002simulation}. While highly accurate, such FEM–DEM coupling is computationally prohibitive for large systems: each particle requires its own finite element mesh, its own elastic solution, and continuous geometry updates or remeshing as contacts evolve. As a result, the computational cost scales with both the number of particles and the particle size, with the latter manifesting through the number of FEM degrees of freedom per particle, rendering such methods impractical for simulations involving more than a modest number of grains. To date, such approaches have not demonstrated scalability to the particle counts typical of granular mechanics, or biological particulate systems. A second class of methods represents each deformable grain as an aggregate of smaller rigid units, typically spheres, connected by springs, thus allowing deformation to arise through their relative motion~\cite{carmona2008fragmentation, andre2012discrete, radi2019elasticity}. They bear similarities to the first family of methods as they effectively discretize each grain, but are relatively simpler to implement. Yet, similarly to the FEM-based methods, these discrete surrogates require a large number of sub-particles to achieve realistic deformation, and their computational cost grows rapidly as a result, rendering them too expensive to scale to systems of practical size. A third class uses modal or reduced-order kinematics in place of a full continuum description, approximating particle deformation through a small number of deformation modes, affine updates, or compliance matrices~\cite{williams1987modal, rojek2018discrete, rojek20213d, mollon2022soft}. These methods are substantially more scalable and can capture certain bulk shape changes, but existing implementations remain limited in several fundamental ways. They typically rely on simplified deformation patterns, are tailored to specific particle shapes, and often lack a transparent connection to fundamental principles of mechanics and thermodynamics. 

In this work, we develop a unified variational formulation for particle deformability derived from first principles and realized through a systematic reduced-order modal description. The approach builds on the observation that classical DEM already resolves \emph{local deformation} at the contacts, and augments this description by incorporating the missing physical ingredient—\emph{bulk deformation}—through a reduced set of modal degrees of freedom. The framework is built upon the same fundamental mechanical structure underlying classical particle simulations, in which rigid-body dynamics is systematically extended in a principled manner through an energetic variational formulation to incorporate particle deformability. In contrast to previous developments, the formulation is not restricted to specific particle geometries and naturally accommodates general deformation modes, as well as nonlinearities in the deformation behavior. It is built on a consistent and extensible theoretical foundation in which all modeling assumptions are explicit. Crucially, the framework is implemented within the Level Set DEM (LS-DEM) formalism~\cite{feldfogel2024discretization, kawamoto2016level}, which naturally accommodates complex particle geometries and topologies, and provides a consistent representation of particle deformation through the evolution of level sets. By integrating the variational formulation with a level set description of grain geometry, the resulting deformable LS-DEM enables a versatile and efficient treatment of evolving complex particle shapes, together with a consistent description of their interactions.

The paper is organized as follows. Section~\ref{theoreticalConstruct} introduces the theoretical framework and presents the detailed variational derivation. Section~\ref{sectionImplementation} describes the numerical implementation within the LS--DEM framework, including the treatment of particle deformation through evolving level sets. Section~\ref{SectionValidation} presents a series of validation examples based on both analytically and numerically obtained deformation modes, and examines the ability of the framework to capture nonlinear behavior. Section~\ref{SectionDiscussion} concludes with a discussion of the main findings and potential extensions.

\section{Theory}
\label{theoreticalConstruct}

\subsection{Preliminaries}
\label{sectionTheoreticalPreliminaries}

In the classical DEM formulation, grains are modeled as rigid bodies whose configuration is fully described by their translational and rotational degrees of freedom; together with their associated rates, these variables define the \emph{state} of the system.  The temporal evolution of this state is governed by a system of ordinary differential equations (Newton’s equations), which may be integrated explicitly in time in order to advance the state of the system. When complemented with appropriate initial conditions, this constitutes a well-posed initial value problem (IVP). 

Extending DEM to account for bulk deformation requires first enriching the description of each grain with additional state variables that encode its evolving shape. In the present formulation, these variables act as generalized deformation coordinates associated with a prescribed set of bulk deformation modes, to be defined rigorously below. They are incorporated together with the translational and rotational degrees of freedom within a single energetic variational framework. This requires extending the energetic structure of the rigid DEM to include, beyond the kinetic and potential energy due to rigid body kinematics, the corresponding kinetic and elastic energy associated with the bulk deformation of the grains. In the presence of non-conservative particle interactions (e.g. friction), a natural framework that accounts for all such energetic modes while consistently accommodating dissipation, is furnished by the d'Alembert principle. The principle states that \textit{among all admissible trajectories, the actual motion of the system is such that the first variation of the conservative action functional is balanced by the virtual work of non-conservative forces}. 

It is worth briefly noting that variational formulations are traditionally formulated as boundary value problems (BVPs), in which the derivation of the governing equations from the underlying variational principle requires the cancellation of temporal boundary terms, which occurs only when variations of the configuration variables are fixed at both ends of the time interval. This appears incompatible with the IVP structure, in which the forward end of time is not known. The standard resolution in analytical mechanics is to first derive the governing differential equations from a formulation with vanishing endpoint variations, and then interpret the resultant equations as the evolution equations of an IVP, which is then supplied with appropriate initial conditions in order to determine the system's trajectory~\cite{rothkopf2023new}. Accordingly, this is the approach we adopt here. The validity of using a boundary-value variational derivation for an IVP has been investigated in both conservative and nonconservative settings~\cite{galley2013classical, bloch1996euler}. In particular, Bloch and coauthors~\cite{bloch1996euler} establish an \emph{if and only if} correspondence between treating the forced Euler–Lagrange equations as the governing evolution equations of dissipative IVPs and the formulation of the Lagrange–d’Alembert variational principle with fixed temporal endpoints.

\subsection{Mathematical Framework}
\label{labelMathematicalDerivation}

Our framework encompasses as state variables the position and rotation vectors, $\vec{X}(t)$ and $\vec{\Theta}(t)$ respectively, which uniquely specify the configuration of each grain at any given time $t$. To incorporate grain deformability, additional state variables $\vec{\varepsilon}(t)$ are introduced that represent internal elastic deformations~\cite{mollon2022soft}. Specifically, the deformation of each grain is expressed as a linear combination of prescribed deformation modes (shape functions) $\vec{\Phi}_{\alpha}$, with associated modal amplitudes $\varepsilon_{\alpha}(t)$, which serve as independent generalized coordinates. The deformation modes define a reduced kinematic basis, in which each mode represents a physically meaningful deformation pattern, and the set of modes is chosen to be sufficiently rich to capture the deformation mechanisms relevant to the problem under consideration.
For the general three-dimensional problem, the complete set of state variables is given by
$\{X_1, X_2, X_3, \Theta_1, \Theta_2, \Theta_3, \varepsilon_1, \varepsilon_2, \dots\}$. 

The local displacement field of a grain due to deformation is given in terms of the deformation degrees of freedom as

\begin{equation}
\vec{u}_{\varepsilon}(\vec{x},t)=\varepsilon_{\alpha}(t)\vec{\Phi}_{\alpha}(\vec{x}),
\label{ResultantDisplacementPerMode}
\end{equation}
where Einstein’s summation convention is used. Throughout the text, repeated Latin indices $i$ denote summation over the three spatial directions associated with the translational and rotational degrees of freedom, $X_i$ and $\Theta_i$. Repeated Greek indices $\alpha,\beta,\gamma$ denote summation over the total number of deformation modes, which may differ from three. The vector field $\vec{\Phi}_{\alpha}(\vec{x})$ is the $\alpha$-th mode shape, and the vector $\vec{x}$ denotes the position in the grain’s local configuration, distinguished from the translational degree of freedom $\vec{X}$.

\begin{figure}[htbp!]
    \centering
    \begin{minipage}{1.0\linewidth}
        \centering
        \includegraphics[width=\linewidth,
                         trim={0cm 0cm 0cm 0cm},clip]
                         {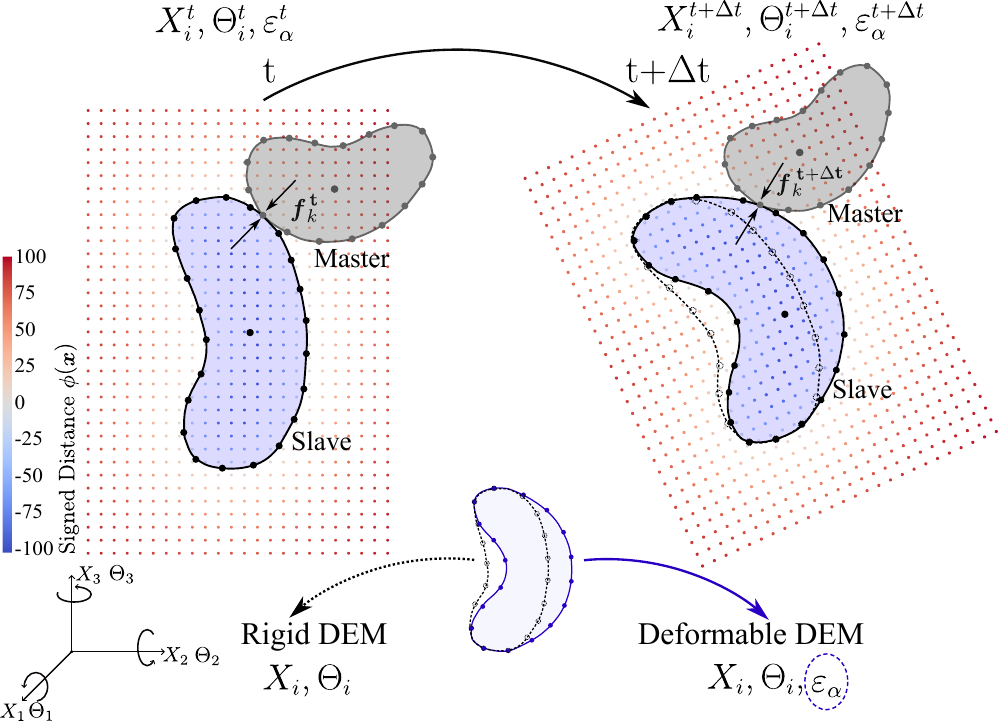}
    \end{minipage}
\caption{
Schematic illustration of the deformable LS--DEM framework. At time $t$, each grain is characterized by its position $X_i^{t}$, orientation $\Theta_i^{t}$, and modal deformation coordinates $\varepsilon_{\alpha}^{t}$, which determine the level set field $\phi^{t}(\boldsymbol{x})$. Interaction forces $f_k^{\,t}$ are computed on the grain surface and drive the evolution of the state variables to $X_i^{t+\Delta t}$, $\Theta_i^{t+\Delta t}$, and $\varepsilon_{\alpha}^{t+\Delta t}$ at time $t+\Delta t$. The right panel shows the updated grain geometry and level set field after deformation. The bottom row highlights the distinction between rigid DEM, where only $X_i$ and $\Theta_i$ evolve, and deformable DEM, where the additional internal deformation coordinates $\varepsilon_{\alpha}$ allow grains to undergo shape changes.}
    \label{fig:GrainStateVariables}
\end{figure}

The form given above constitutes a \emph{reduced} kinematic description, in which the displacement field is approximated through a separation of spatial and temporal dependencies. In this formulation, the mode shapes are assumed to be fixed in time; that is, the spatial distribution of each deformation mode remains unchanged, while its associated generalized coordinate evolves in time. Consequently, the shape functions $\vec{\Phi}_{\alpha}(\vec{x})$ are independent of both the generalized coordinates $\varepsilon_{\alpha}$ and time. The displacement field is thus represented as a finite sum of fixed spatial shape functions with time-dependent amplitudes. Specific strategies for constructing an efficient basis of modes in a tailored problem-specific manner are discussed in Section~\ref{SectionValidation}.

It is worth emphasizing that the use of fixed mode shapes does not restrict the formulation to an infinitesimal-strain regime. The present assumption concerns only the choice of a fixed modal basis, which can accommodate large strains through large values of the generalized coordinates $\varepsilon_{\alpha}$, provided that the dominant deformation mechanisms are reasonably represented by the chosen fixed shape functions. Throughout the following derivation, the formulation is carried out without invoking a small-strain or linearized constitutive assumption; such a simplification is introduced only at the final stage in deriving the governing equations. This preserves a clear structure amenable to subsequent extensions discussed in more detail in Section~\ref{SectionDiscussion}.

To embed this enriched description into a variational energetic framework based on the Lagrange--d'Alembert principle, we must specify the energetic contributions associated with the additional state variables describing deformation. In particular, this entails deriving explicit expressions for the elastic potential energy and the kinetic energy associated with deformation, written in terms of the generalized coordinates $\varepsilon_{\alpha}$.
In the special case of linear elastic material behavior, using the displacement representation given in Eq.~\ref{ResultantDisplacementPerMode} together with the linearity of the infinitesimal strain operator, the corresponding strain field reads
$\boldsymbol{E}(\mathbf{u}) = \varepsilon_{\alpha}\,\boldsymbol{E}(\boldsymbol{\Phi}_{\alpha})$, while the constitutive law reads $\boldsymbol{\sigma}=\mathbb{C}:\boldsymbol{E}$. Hence, the elastic energy is given by
\begin{equation}
PE_e = \frac{1}{2}\int_\Omega \boldsymbol{\sigma}:\boldsymbol{E}\,\mathrm{d}V,
\label{linearElasticPotentialEnergy}
\end{equation}
which, for constant shape functions, reduces to the quadratic form
\begin{equation}
PE_e = \frac{1}{2}K_{\alpha\beta}\,\varepsilon_{\alpha}\,\varepsilon_{\beta},
\label{ResultantEnergyLinear}
\end{equation}
where the modal stiffness coefficients $K_{\alpha\beta}$—which constitute the generalized stiffness matrix of the grain—are given by
\begin{equation}
K_{\alpha\beta} = \int_{\Omega} \bigl(\mathbb{C}:\boldsymbol{E}(\boldsymbol{\Phi}_{\alpha})\bigr) :
\boldsymbol{E}(\boldsymbol{\Phi}_{\beta})\,\mathrm{d}V,
\label{ModalStiffnessDefinition}
\end{equation}
where $\Omega$ denotes the grain’s reference configuration, and $\boldsymbol{E}(\boldsymbol{\Phi}_{\alpha})$ is the strain field induced by the $\alpha$-th deformation mode. 

The expressions above are derived under the assumption of linear elastic constitutive behavior and small strain measure, together with fixed deformation mode shapes, which leads to a quadratic elastic potential energy in terms of the generalized deformation coordinates. This corresponds to the case of deformation modes governed by linear force–displacement relations, referred to as linear modes. However, the reduced kinematic description adopted in this work---based on constant shape functions---does not inherently restrict the elastic potential energy to be quadratic, as it is not intrinsically tied to small-strain kinematics or linear elastic material behavior. The reduced variational structure introduced in this work is compatible with both infinitesimal- and finite-strain formulations. Reduced elastic energies in finite-strain settings, including nonlinear constitutive behavior, may be constructed in terms of the generalized coordinates $\varepsilon_{\alpha}$ by adopting finite-strain kinematics and appropriate hyperelastic energy densities, leading to nonlinear generalized force-displacement relationship at the modal level.

Such nonlinear modal responses may also arise in the absence of intrinsically nonlinear material behavior. In particular, systems composed of linearly elastic particles operating within the small-strain regime may nevertheless exhibit nonlinear reduced-order dynamics due to geometric and interaction-driven effects. For instance, evolving contact configurations can give rise to nonlinear force – displacement relations at the level of the generalized modal coordinates. A detailed discussion of the nature of these modal nonlinearities, together with their systematic incorporation within the proposed framework, is provided in Section~\ref{NumericallyDerivedModesSection}.

Accordingly, we allow the potential energy associated with the set of deformation modes to be represented by an arbitrary, possibly non-quadratic, energy functional $U_\varepsilon(\varepsilon_{\alpha})$ of the modal amplitudes, in order to accommodate nonlinear relations between generalized modal forces and displacements. Here, $\varepsilon_{\alpha}$ denotes the complete set of modal amplitudes upon which the energy functional depends. To derive the kinetic energy contribution associated with bulk deformation~\cite{mollon2022soft}, we begin by differentiating the modal displacement representation in Eq.~\ref{ResultantDisplacementPerMode}:
\begin{equation}
\dot{\mathbf{u}}( \vec{x},t) = \dot{\varepsilon}_{\alpha}(t)\,\boldsymbol{\Phi}_{\alpha}( \vec{x}).
\end{equation}
Let $\rho( \vec{x})$ denote the mass density in the grain’s reference configuration $\Omega$. The kinetic energy due to deformation is then
\begin{equation}
KE = \frac{1}{2}\int_{\Omega} \rho( \vec{x})\,\dot{\mathbf{u}}( \vec{x},t)\cdot \dot{\mathbf{u}}( \vec{x},t)\,\mathrm{d}V.
\end{equation}
Substituting $\dot{\mathbf{u}} = \dot{\varepsilon}_{\alpha}\,\boldsymbol{\Phi}_{\alpha}$ yields
\begin{align}
KE
&= \frac{1}{2}\dot{\varepsilon}_{\alpha}\,\dot{\varepsilon}_{\beta}
\int_{\Omega} \rho( \vec{x})\,\boldsymbol{\Phi}_{\alpha}( \vec{x})\cdot\boldsymbol{\Phi}_{\beta}( \vec{x})\,\mathrm{d}V.
\end{align}
This motivates the definition of the generalized (modal) mass matrix
\begin{equation}
M_{\alpha\beta} = \int_{\Omega} \rho( \vec{x})\,\boldsymbol{\Phi}_{\alpha}( \vec{x})\cdot \boldsymbol{\Phi}_{\beta}( \vec{x})\,\mathrm{d}V,
\label{ModalMassDefinition}
\end{equation}
so that the kinetic energy associated with deformation takes the compact form
\begin{equation}
KE = \frac{1}{2} M_{\alpha\beta}\,\dot{\varepsilon}_{\alpha}\,\dot{\varepsilon}_{\beta}.
\label{ResultantKineticGeneral}
\end{equation}

In contrast to the elastic potential energy term given in Eq.~\ref{ResultantEnergyLinear}, the kinetic energy associated with the deformation modes is of purely kinematic origin and does not depend on the constitutive force–displacement behavior of the modes. Assuming fixed deformation shape functions, its functional form follows uniquely from the chosen reduced kinematic description. As a result, the expression given in Eq.~\ref{ResultantKineticGeneral} may be incorporated directly into the variational formulation, independently of whether the response is linear or nonlinear.

Having introduced the abstract framework for incorporating additional degrees of freedom to represent elastic deformations, together with their corresponding energy expressions in terms of the generalized coordinates, we can begin formulating the \emph{Lagrange--d’Alembert principle}, whose mathematical representation is

\begin{equation}
    \delta \int_a^b L\big(q(t),\dot{q}(t)\big)\, dt 
+ \int_a^b F\big(q(t),\dot{q}(t)\big) \cdot \delta q \, dt = 0.
\label{FundPrinciple}
\end{equation}
The first term is the first variation of the action of the system, often denoted by $\delta S$, while the second term is the virtual work of the imposed forces, often denoted by $\delta W$. This principle provides the foundation for deriving the governing equations of motion in terms of the chosen generalized coordinates, harnessing variational principles and energetic arguments, while consistently incorporating dissipative effects.

At the level of a single particle (grain), the statement of the Lagrange--d’Alembert principle consists of the first variation of the action, constructed from the kinetic and elastic potential energies of the grain, together with the virtual work of external forces acting on the grain.  

In the present formulation, assuming that the particle surface is discretized into nodes, as in the LS--DEM framework (see Section~\ref{sectionImplementation}), this virtual work contribution is written in terms of discrete forces associated with surface nodes. This reflects the numerical representation adopted here, in which forces and displacements are defined only at nodal points. This yields the fundamental balance
\begin{gather}
    \delta \left[ 
\int_{t_1}^{t_2} \Big( KE(\dot{X}_i,\dot{\Theta}_i,\dot{\varepsilon}_{\alpha}) 
- PE_e(\varepsilon_{\alpha})  \Big) \,\mathrm{d}t \right] \nonumber\\
+ \int_{t_1}^{t_2} \Big( \sum_{k=1}^{n}  \boldsymbol{f}_k(X_i,\Theta_i,\varepsilon_{\alpha}) \ \cdot \ \delta\left[  \boldsymbol{u}_k (X_i, \Theta_i, \varepsilon_{\alpha})\right] \Big) \,\mathrm{d}t = 0.
\end{gather}
In the second term, the summation over $k$ runs over the surface nodes of the grain, with $\boldsymbol{f}_k$ denoting the force acting on node $k$ and $\boldsymbol{u}_k$ the corresponding nodal displacement. This equation represents the fundamental variational statement governing the motion of a deformable grain, expressed in a form consistent with the LS--DEM framework. The variational statement naturally extends to a system of particles by summing over all grains, where nonconservative forces arise from particle interactions (e.g., friction).

The next step is to follow the usual variational procedure which involves perturbing each generalized coordinate—translation, rotation, and modal amplitude—according to $q \mapsto q + \epsilon\,\eta_q$, where $\eta_q$ denotes an admissible variation, and differentiating with respect to $\epsilon$ at $\epsilon=0$. We first apply this procedure to the virtual work term (Section~\ref{sectionVariationVirtualWork}), which yields the functional form of the generalized forces conjugate to the translational, rotational, and deformation coordinates. We then apply the same procedure to the action term (Section~\ref{SectionVariationAction}), and finally combine the two contributions in a single integral to obtain the final explicit form of the governing equations.

\subsubsection{Variation of the virtual work}
\label{sectionVariationVirtualWork}

The following variations are introduced:
\begin{equation}
X_i \mapsto X_i + \epsilon\,\eta_{X_i},
\end{equation}
\begin{equation}
\Theta_i \mapsto \Theta_i + \epsilon\,\eta_{\Theta_i},
\end{equation}
\begin{equation}
\varepsilon_{\alpha} \mapsto \varepsilon_{\alpha} + \epsilon\,\eta_{\varepsilon_{\alpha}},
\end{equation}
\begin{equation}
\dot{X}_i \mapsto \dot{X}_i + \epsilon\,\dot{\eta}_{X_i},
\end{equation}
\begin{equation}
\dot{\Theta}_i \mapsto \dot{\Theta}_i + \epsilon\,\dot{\eta}_{\Theta_i},
\end{equation}
\begin{equation}
\dot{\varepsilon}_{\alpha} \mapsto \dot{\varepsilon}_{\alpha} + \epsilon\,\dot{\eta}_{\varepsilon_{\alpha}}.
\end{equation}

The virtual work $\delta W$ follows as
\begin{equation}
\delta W =
\int_{t_1}^{t_2}
\hspace{-0.1cm}\big(
\sum_{k=1}^{n}
\boldsymbol{f}_k(X_i,\Theta_i,\varepsilon_{\alpha})\!\cdot\!
\left.
\frac{\mathrm{d}}{\mathrm{d}\epsilon}\,
\boldsymbol{u}_k\big(
X_i\!+\!\epsilon\eta_{X_i},
\Theta_i\!+\!\epsilon\eta_{\Theta_i},
\varepsilon_{\alpha}\!+\!\epsilon\eta_{\varepsilon_{\alpha}}
\big)
\right|_{\epsilon = 0}
\big)\,\mathrm{d}t.
\end{equation}

By the chain rule, we obtain
\begin{equation}
   \delta W =  \int_{t_1}^{t_2}  \big(  \sum_{k=1}^{n} \boldsymbol{f}_k(X_i,\Theta_i,\varepsilon_{\alpha}) \ \cdot ( \frac{\partial \boldsymbol{u}_k }{\partial X_i} \eta_{X_i} + \frac{\partial \boldsymbol{u}_k }{\partial \Theta_i} \eta_{\Theta_i}+ \frac{\partial \boldsymbol{u}_k}{\partial \varepsilon_{\alpha} }\eta_{\varepsilon_{\alpha}} ) \big) \mathrm{d}t,
   \label{virtualWorkFund}
\end{equation}
where the partial derivative of the displacement vector at a node is

\begin{equation}
\frac{\partial \boldsymbol{u}_k }{\partial X_i} = \scalebox{1.2}{$\boldsymbol{e}_{i_k}$},
\end{equation}
where $\boldsymbol{e}_{i_k}$ is the unit vector in the direction of $X_i$ evaluated at node $k$. The translational degrees of freedom generate uniform displacements throughout the grain. Consequently, the change in the displacement vector of any node with respect to a unit variation of a translational degree of freedom is independent of the grain position, orientation, shape, or the current location of the node, and is given by the fixed unit vector in the corresponding direction.

For the rotational degrees of freedom, we obtain:
\begin{equation}
\frac{\partial \boldsymbol{u}_k }{\partial \Theta_i} = r_k(\varepsilon_{\alpha})  \scalebox{1.2}{$\boldsymbol{e}_{\theta_i}$}_k(\Theta_i,\varepsilon_{\alpha}),
\end{equation}
where \( r_k(\varepsilon_{\alpha}) \) is the current (at time \(t\)) distance of the lever arm of the node $k$, which is a function of the deformation and therefore expressed as \( r_k(\varepsilon_{\alpha}) \). The second term, \( \boldsymbol{e}_{\theta_i}(\Theta_i,\varepsilon_{\alpha}) \), is the current unit vector in the \(\Theta_i\) direction of the node $k$, which evolves with both grain rotation and deformation and therefore depends on the rotational and deformation degrees of freedom.

Finally, for the deformation degrees of freedom we have 
\begin{equation}
\frac{\partial \boldsymbol{u}_k }{\partial \varepsilon_{\alpha}} = {\Phi_{\alpha}}_k \scalebox{1.2}{$\boldsymbol{e}_{\varepsilon_{\alpha}}$}_k(\Theta_i),
\end{equation}
where \( \boldsymbol{e}_{\varepsilon_{\alpha} k}(\Theta_i) \) is the unit vector in the direction of the displacement induced by deformation, and ${\Phi_{\alpha}}_k$ is the value of the shape function $\vec{\Phi}_{\alpha}$ evaluated at node $k$. This direction is determined by the prescribed shape function and remains constant in the local frame of the grain even in the presence of deformation due to the fixed shape functions used in our reduced kinematic description. Thus, it does not depend on the deformation degrees of freedom. However, as the grain rotates, this direction rotates in space and therefore depends on the rotational degrees of freedom $\Theta_i$.

Inserting these expressions into Eq.~\ref{virtualWorkFund}, we obtain 
\begin{gather}
\delta W =
\int_{t_1}^{t_2}
\Big(
\sum_{k=1}^{n}
\boldsymbol{f}_k(X_i,\Theta_i,\varepsilon_{\alpha})
\ \cdot \big(
\scalebox{1.2}{$\boldsymbol{e}_{i_k}$}\eta_{X_i} + r_k(\varepsilon_{\alpha})\,\scalebox{1.2}{$\boldsymbol{e}_{\theta_i}$}_k(\Theta_i,\varepsilon_{\alpha})\eta_{\Theta_i}
\nonumber\\
\qquad\qquad
+ \ {\Phi_{\alpha}}_k\,\scalebox{1.2}{$\boldsymbol{e}_{\varepsilon_{\alpha}}$}_k(\Theta_i)\eta_{\varepsilon_{\alpha}}
\big)
\Big)\,\mathrm{d}t .
\end{gather}

We can move the dot product and the summation inside the parentheses, dropping the explicit functional dependence of $\boldsymbol{f}_k$ for conciseness.
\begin{gather}
\delta W =
\int_{t_1}^{t_2} \Big(
\sum_{k=1}^{n} \boldsymbol{f_k} \cdot \scalebox{1.2}{$\vec{e}_{i_k}$}\,\eta_{X_i}
+ \sum_{k=1}^{n} r_k(\varepsilon_{\alpha})\,\boldsymbol{f_k} \cdot
  \scalebox{1.2}{$\vec{e}_{\theta_i}$}_k (\Theta_i,\varepsilon_{\alpha})\,\eta_{\Theta_i}
\nonumber\\[4pt]
\qquad\qquad
+ \sum_{k=1}^{n} \Phi_{\alpha k}\,\boldsymbol{f_k} \cdot
  \scalebox{1.2}{$\vec{e}_{\varepsilon_{\alpha}}$}_k (\Theta_i)\,\eta_{\varepsilon_{\alpha}}
\Big)\, \mathrm{d}t.
\end{gather}

This expression is now written in the canonical variational form, where each variation of a generalized degree of freedom is multiplied by its corresponding \textit{conjugate} generalized force.
\begin{equation}
   \delta W = \int_{t_1}^{t_2} \big( F_{i}  \eta_{X_i} + \scalebox{1.2}{$\tau_i$}\eta_{\Theta_i} + F_{\varepsilon_{\alpha}} \eta_{\varepsilon_{\alpha}} \big) \mathrm{d}t,
   \label{VirtualWorkTerm}
\end{equation}
where the generalized forces are given by
\begin{equation}
F_i =  \sum_{k=1}^{n}  \ \boldsymbol{f}_k \cdot \scalebox{1.2}{$\boldsymbol{e}_{i_k}$},
\label{GF1}
\end{equation}

\begin{equation}
\scalebox{1.2}{$\tau_i$} = \sum_{k=1}^{n} r_k(\varepsilon_{\alpha}) \ \boldsymbol{f}_k \cdot  \scalebox{1.2}{$\boldsymbol{e}_{\theta_i}$}_k (\Theta_i,\varepsilon_{\alpha}),
\label{GF2}
\end{equation}

\begin{equation}
F_{\varepsilon_\alpha}=  \sum_{k=1}^{n} {\Phi_{\alpha}}_k  \boldsymbol{f}_k \cdot \scalebox{1.2}{$\boldsymbol{e}_{\varepsilon_{\alpha}}$}_k(\Theta_i).
\label{GF3}
\end{equation}

\subsubsection{Variation of the action}
\label{SectionVariationAction}
We now follow the same variational procedure for the first term in the Lagrange--d’Alembert statement, namely the first variation of the action. The Lagrangian is defined as the difference between kinetic and potential energies. In addition to the translational and rotational contributions, the kinetic energy now also includes the deformation-related term introduced earlier. Its functional form is given by
\begin{equation}
KE(t) = \tfrac{1}{2} M_i \dot{X}_i^2 \;+\; \tfrac{1}{2} I_i(\varepsilon_{\gamma})\, \dot{\Theta}_i^2 \;+\; \tfrac{1}{2} M_{\alpha\beta}(\varepsilon_{\gamma})\, \dot{\varepsilon}_{\alpha} \dot{\varepsilon}_{\beta},
\end{equation}
where $M_i$ and $I_i$ denote the mass and the principal moments of inertia of the grain, respectively, and $M_{\alpha\beta}$ are the components of the generalized mass matrix defined in Eq.~\ref{ModalMassDefinition}. The subscript $\gamma$ in the arguments of $I_i(\varepsilon_{\gamma})$ and $M_{\alpha\beta}(\varepsilon_{\gamma})$ indicates that these quantities generally depend on the full set of deformation coordinates, following the same notational convention as for the elastic potential energy. With these definitions in place, the first variation of the action\footnote{A gravitational potential term, $f_G = -M_i g,X_{3}$ (or its equivalent in the chosen coordinate system), can be added straightforwardly to the Lagrangian. It enters simply as an additional potential energy term and does not alter the variational structure.} can be written explicitly as shown below, where the elastic potential is represented by the general function $U_\varepsilon(\varepsilon_{\alpha})$, to be later specialized to the case of linear elasticity.
\begin{equation}
\hspace{-0.35cm}\delta S
= \delta \left[ 
\int_{t_1}^{t_2} 
\left( 
\frac{1}{2} M_i \,\dot{X}_i^{\,2}
\!+\!\frac{1}{2} I_i(\varepsilon_\gamma) \,\dot{\Theta}_i^{\,2}
\!+\!\frac{1}{2} M_{\alpha\beta}(\varepsilon_\gamma)\, \dot{\varepsilon}_\alpha \dot{\varepsilon}_\beta
\!-\!U_\varepsilon(\varepsilon_\alpha)
\right) \mathrm{d}t 
\right].
\label{ActionFunctionalFormVersion}
\end{equation}

Introducing the variations defined earlier for the degrees of freedom, the above can be rewritten as
\begin{gather}
\hspace{-1.0cm}\delta S =
\frac{d}{d\epsilon} \Bigg[
\int_{t_1}^{t_2} \Big(
\tfrac{1}{2} M_i (\dot{X}_i + \dot{\epsilon \eta_{x_i}} )^2
+ \tfrac{1}{2} I_i(\varepsilon_{\gamma} + \epsilon \eta_{\varepsilon_{\gamma}})
  (\dot{\Theta}_i + \dot{\epsilon \eta_{\Theta_i}} )^2
 \nonumber\\[4pt]
\hspace{-1.25cm}\qquad\qquad
+ \tfrac{1}{2} M_{\alpha\beta}(\varepsilon_{\gamma} + \epsilon \eta_{\varepsilon_{\gamma}})
  (\dot{\varepsilon}_{\alpha} + \dot{\epsilon \eta_{\varepsilon_{\alpha}}})
  (\dot{\varepsilon}_{\beta} + \dot{\epsilon \eta_{\varepsilon_{\beta}}})
- U_\varepsilon(\varepsilon_{\alpha} + \epsilon \eta_{\varepsilon_{\alpha}})
\Big) \,\mathrm{d}t \Bigg]_{\epsilon = 0}.
\label{StationaryConditionEpsilonDerivative}
\end{gather}

Since the variation parameter $\epsilon$ and time $t$ are independent variables, differentiation with respect to $\epsilon$ commutes with time integration. The first variation of the action is therefore obtained by bringing the derivative inside the integral, expanding the resulting expression in terms of the generalized coordinates and their rates, enforcing $\epsilon=0$, and applying integration by parts to transfer time derivatives from the variations to the generalized coordinates. Then, based on the argumentation provided in Section~\ref{sectionTheoreticalPreliminaries}, admissible variations are enforced to vanish at the temporal boundaries, i.e.\ at $t=t_1$ and $t=t_2$, which eliminates the boundary terms arising from integration by parts.  The complete derivation, including the intermediate $\epsilon$-dependent expressions, is provided in~\ref{appendixVariationalDerivation}. These steps yield the following form of the stationarity condition:
\begin{gather}
\delta S =
\int_{t_1}^{t_2} \Big(
- M_i \ddot{X}_i\, \eta_{X_i}
+ \tfrac{1}{2} I_i'(\varepsilon_{\gamma})\, \eta_{\varepsilon_{\gamma}}\, \dot{\Theta}_i^{\,2}
- I_i(\varepsilon_{\gamma})\, \ddot{\Theta}_i\, \eta_{\Theta_i}
- I_i'(\varepsilon_{\gamma})\, \dot{\varepsilon}_{\gamma}\, \dot{\Theta}_i\, \eta_{\Theta_i}
 \nonumber\\
\qquad\qquad
+\tfrac{1}{2} M_{\alpha\beta}'(\varepsilon_{\gamma})\, \eta_{\varepsilon_{\gamma}}\, \dot{\varepsilon}_{\alpha} \dot{\varepsilon}_{\beta}
- M_{\alpha\beta}(\varepsilon_{\gamma})\, \ddot{\varepsilon}_{\beta}\, \eta_{\varepsilon_{\alpha}}
- M_{\alpha\beta}'(\varepsilon_{\gamma})\, \dot{\varepsilon}_{\gamma} \dot{\varepsilon}_{\alpha}\, \eta_{\varepsilon_{\alpha}} \nonumber\\
\qquad\qquad
- U_\varepsilon'(\varepsilon_{\alpha})\, \eta_{\varepsilon_{\alpha}}
\Big)\,\mathrm{d}t.
\label{ExpressionDerivedInAppendix}
\end{gather}

Here, the prime notation denotes differentiation of a function with respect to its argument, evaluated in the direction of the admissible variation. This convention applies to $I_i(\varepsilon_\gamma)$, $M_{\alpha\beta}(\varepsilon_\gamma)$, and $U_\varepsilon(\varepsilon_\alpha)$.

Henceforth, the deformation-dependence of the moments of inertia and of the generalized mass matrix is neglected for the sake of conciseness, following a small deformation assumption. Accordingly, the inertia quantities are treated as constants and all terms involving their derivatives with respect to $\varepsilon_\alpha$ are dropped\footnote{Retaining the deformation-dependence of the inertia quantities would extend the present theoretical framework to large-deformation settings and would lead, through the continuation of the same variational derivation, to additional inertia-related contributions in the governing equations.}. Under this assumption, the first variation of the action reduces to
\begin{equation}
\delta S =
\int_{t_1}^{t_2} \Big(
- M_i\, \ddot{X}_i\, \eta_{X_i}
- I_i\, \ddot{\Theta}_i\, \eta_{\Theta_i}
- M_{\alpha\beta}\, \ddot{\varepsilon}_{\beta}\, \eta_{\varepsilon_{\alpha}}
- U_\varepsilon'(\varepsilon_{\alpha})\, \eta_{\varepsilon_{\alpha}}
\Big)\,\mathrm{d}t,
\label{FirstVariationAction}
\end{equation}
which completes the derivation of the first variation of the action in terms of the generalized degrees of freedom and their variations.

\subsubsection{Governing equations}
\label{SectionGoverningEquations}
We then combine the two contributions—the first variation of the action (Eq.~\ref{FirstVariationAction}) and the virtual work (Eq.~\ref{VirtualWorkTerm})—into a single integral
\begin{gather}
\int_{t_1}^{t_2} \Big(
- M_i\, \ddot{X}_i\, \eta_{X_i}
- I_i\, \ddot{\Theta}_i\, \eta_{\Theta_i}
- M_{\alpha\beta}\, \ddot{\varepsilon}_{\beta}\, \eta_{\varepsilon_{\alpha}}
- U_\varepsilon'(\varepsilon_{\alpha})\, \eta_{\varepsilon_{\alpha}}
\nonumber\\[4pt]
\qquad\qquad\qquad
+ F_i\, \eta_{X_i}
+ \scalebox{1.2}{$\tau_i$}(\Theta_i,\varepsilon_{\alpha})\, \eta_{\Theta_i}
+ F_{\varepsilon_{\alpha}}(\Theta_i)\, \eta_{\varepsilon_{\alpha}}
\Big)\,\mathrm{d}t = 0.
\label{StationaryCondition}
\end{gather}
Rearranging the terms that are multiplying the same variations
\begin{gather}
\int_{t_1}^{t_2} \Big(
\big(F_i - M_i \ddot{X}_i\big)\, \eta_{X_i}
+ \big(\scalebox{1.2}{$\tau_i$}(\Theta_i,\varepsilon_{\alpha}) - I_i \ddot{\Theta}_i\big)\, \eta_{\Theta_i}
\nonumber\\[4pt]
\qquad\qquad\qquad
+ \big(F_{\varepsilon_{\alpha}}(\Theta_i) - U_{E_j}'(\varepsilon_{\alpha}) - M_{\alpha\beta}\, \ddot{\varepsilon}_{\beta}\big)\, \eta_{\varepsilon_{\alpha}}
\Big)\,\mathrm{d}t = 0.
\end{gather}

For the integral to vanish, each term multiplying each variation must vanish. This yields the following set of equations:
\begin{equation}
M_i\, \ddot{X}_i = F_i,
\label{FinalSetOfEquationsAbstract1} 
\end{equation}
\begin{equation}
I_i\, \ddot{\Theta}_i = \scalebox{1.2}{$\tau_i$}(\Theta_i,\varepsilon_{\alpha}),
\label{FinalSetOfEquationsAbstract2} 
\end{equation}
\begin{equation}
M_{\alpha\beta}\, \ddot{\varepsilon}_{\beta} + U_\varepsilon'(\varepsilon_{\alpha}) = F_{\varepsilon_{\alpha}}(\Theta_i).
\label{FinalSetOfEquationsAbstract3} 
\end{equation}

In the general case, the governing equations involve the derivative of the elastic potential, $U_\varepsilon'(\varepsilon_{\alpha})$, which is identified with the \emph{internal} generalized force,
$F^{\mathrm{int}}_{\varepsilon_\alpha} = U_\varepsilon'(\varepsilon_{\alpha})$.
The governing equations~\eqref{FinalSetOfEquationsAbstract1}--\eqref{FinalSetOfEquationsAbstract3} therefore remain valid for arbitrary, possibly nonlinear, modal responses, provided that an appropriate energy functional $U_\varepsilon$ is specified. In this setting, the reduced kinematic description with fixed shape functions remains unchanged, while nonlinear material and geometric effects are incorporated exclusively through the choice of the modal energy functional. A concrete construction of nonlinear effective modal energies $U_\varepsilon$, together with the corresponding theoretical considerations, is presented in Section~\ref{NumericallyDerivedModesSection}, along with a representative application example. In the linear-elastic case, the elastic potential reduces to $U_\varepsilon = \tfrac{1}{2} K_{\alpha\beta}\,\varepsilon_{\alpha} \,\varepsilon_{\beta}$ (see Eq.~\ref{ResultantEnergyLinear}), whose derivative is $U_\varepsilon'(\varepsilon_{\alpha}) = K_{\alpha\beta}\,\varepsilon_{\beta}$. Substituting this expression, together with the expressions derived for the generalized external forces (Eqs.~\ref{GF1}-\ref{GF3}), yields the final component form of the evolution equations for the linear setting,
\begin{equation}
    M_i \ddot{X}_i =  \sum_{k=1}^{n} \boldsymbol{f}_k\cdot
    \scalebox{1.2}{$\boldsymbol{e}_{i_k}$},
\label{FinalSetOfEquationsDetailed1}
\end{equation}
\begin{equation}
I_i \ddot{\Theta}_i = \sum_{k=1}^{n} r_k(\varepsilon_{\alpha})\,
    \boldsymbol{f}_k\cdot
    \scalebox{1.2}{$\boldsymbol{e}_{\theta_i}$}_k(\Theta_i,\varepsilon_{\alpha}),
\label{FinalSetOfEquationsDetailed2}
\end{equation}
\begin{equation}
    M_{\alpha\beta}\, \ddot{\varepsilon}_{\beta} + K_{\alpha\beta}\,\varepsilon_{\beta} =
    \sum_{k=1}^{n} {\Phi_{\alpha}}_k\, \boldsymbol{f}_k\cdot
    \scalebox{1.2}{$\boldsymbol{e}_{\varepsilon_{\alpha}}$}_k(\Theta_i).
\label{FinalSetOfEquationsDetailed3}
\end{equation}
In tensorial form, the translational, rotational, and modal subsystems can be expressed compactly using direct notation. Introducing the vectors $ \vec{X} = (X_1,X_2,X_3)^{\mathsf T}$, $\boldsymbol{\Theta} = (\Theta_1,\Theta_2,\Theta_3)^{\mathsf T}$, and $\boldsymbol{\varepsilon} = (\varepsilon_1,\ldots,\varepsilon_n)^{\mathsf T}$, together with the diagonal inertia matrices $\mathbf{M}_X = \operatorname{diag}(M_1,M_2,M_3)$ and $\mathbf{I}_\Theta = \operatorname{diag}(I_1,I_2,I_3)$, and the modal mass and stiffness matrices
$\mathbf{M}_{\varepsilon} = [M_{\alpha\beta}]$ and $\mathbf{K}_{\varepsilon} = [K_{\alpha\beta}]$, the governing equations take the form
\begin{equation}
\mathbf{M}_X\,\ddot{ \vec{X}} = \mathbf{F}
\label{linearTensorialEquations1}
\end{equation}
\begin{equation}
    \mathbf{I}_\Theta\,\ddot{\boldsymbol{\Theta}} = \boldsymbol{\tau},
\label{linearTensorialEquations2}
\end{equation}
\begin{equation}
    \mathbf{M}_\varepsilon\,\ddot{\boldsymbol{\varepsilon}}
    + \mathbf{K}_\varepsilon\,\boldsymbol{\varepsilon}
    = \mathbf{F}_\varepsilon.
\label{linearTensorialEquations3}
\end{equation}

The direct tensorial form makes the physical structure of the modal subsystem particularly transparent. The governing equations define a reduced-order dynamical system that couples translational motion, rigid-body rotation, and internal grain deformation within a unified framework. In particular, the equation $\mathbf{M}_\varepsilon\,\ddot{\boldsymbol{\varepsilon}} + \mathbf{K}_\varepsilon\,\boldsymbol{\varepsilon} = \mathbf{F}_\varepsilon
$ reveals that the elastic degrees of freedom behave as a high-dimensional generalized mass--spring system, in which each modal coordinate carries an effective inertia and stiffness, while coupling between modes is encoded in the off-diagonal entries of $\mathbf{M}_\varepsilon$ and $\mathbf{K}_\varepsilon$. For the special case in which the deformation modes are orthogonal with respect to both the generalized mass and stiffness inner products (Eqs.~\ref{ModalMassDefinition} and~\ref{ModalStiffnessDefinition}), the matrices $\mathbf{M}_\varepsilon$ and $\mathbf{K}_\varepsilon$ become diagonal. In this setting, the elastic subsystem reduces to a set of uncoupled second-order ordinary differential equations,
\[
m_{\alpha}\,\ddot{\varepsilon}_{\alpha} + k_{\alpha}\,\varepsilon_{\alpha}
= F_{\varepsilon_{\alpha}}, \qquad \text{with no summation over repeated index }\alpha,
\]
so that each modal coordinate evolves independently of the others. 

Importantly, the formulation is expressed entirely in terms of ordinary differential equations involving a finite number of generalized coordinates per particle, thereby preserving the computational structure of classical DEM: each additional deformation mode introduces one additional scalar evolution equation to the standard set of equations associated with the translational and rotational degrees of freedom of rigid-body DEM.

At this stage, the framework remains agnostic to the specific choice of deformation modes, the numerical representation of particle geometry, and the detailed evaluation of contact and bonding forces. These aspects are addressed in the following section.

\section{Numerical Implementation within LS-DEM}
\label{sectionImplementation}

This section describes the numerical implementation of deformable LS--DEM within the
original LS--DEM framework~\cite{kawamoto2016level,feldfogel2024discretization}, where the particle geometry is represented by the surface nodes and a level set discretized on a regular grid (Fig.~\ref{fig:GrainStateVariables}), and contact is detected via a surface node-to-level set algorithm. Hence in the proposed deformable LS-DEM framework, both the surface nodes and level set must be updated to account for particle deformation. The nodal coordinates of the surface mesh are updated directly according to the shape functions of the deformation modes and the current values of the associated generalized degrees of freedom, as defined in Eq.~\ref{ResultantDisplacementPerMode}. The more challenging task is to efficiently update the level set function based on the deformed surface mesh, preserving accuracy in the vicinity of the particle boundary (i.e., zero level set), thereby ensuring  accurate interparticle force computations.
The numerical implementation also encompasses the evaluation of contact and bonding interactions using the evolving geometry,  the computation of generalized forces associated with the deformation modes, and the time integration of the resulting equations of motion. 

\subsection{Deformation of the level set}
\begin{figure}[htbp!]
    \centering
    \begin{minipage}{0.9\linewidth}
        \centering
        \includegraphics[width=\linewidth,
                         trim={0cm 0cm 0cm 0cm},clip]
                         {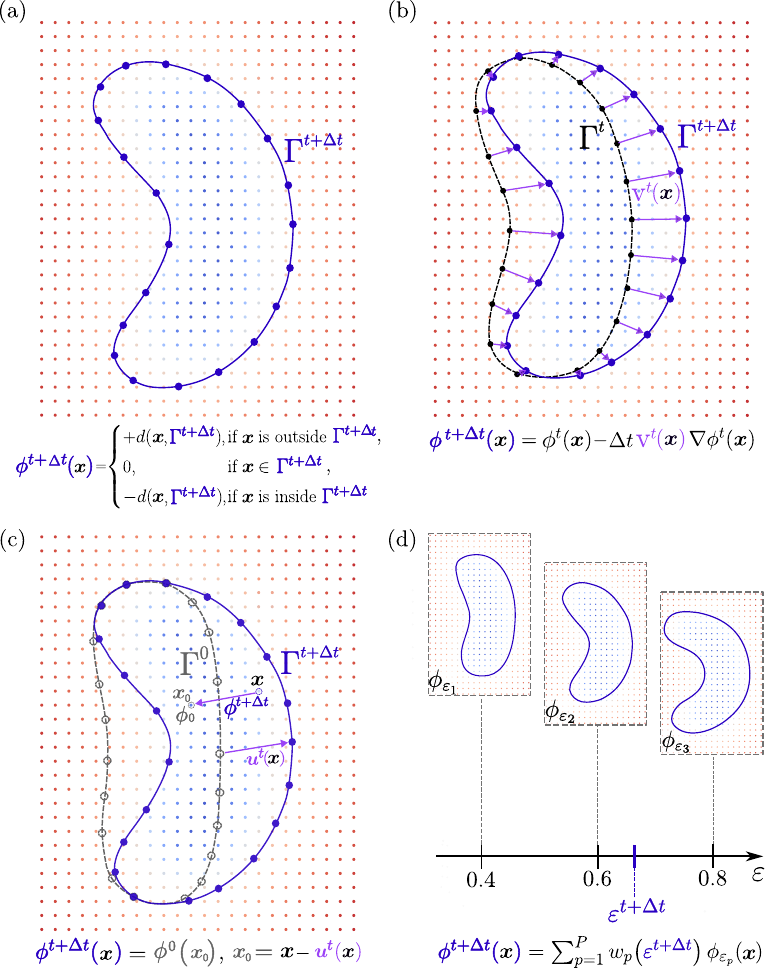}
    \end{minipage}

\caption{
Illustration of several possible approaches for updating the level set field under 
particle deformation (shown in 2D for ease of illustration).  
(a) Direct recomputation of the level set field from the updated particle 
surface at time $t+\Delta t$.  
(b) Advection of the existing level set field by solving a transport equation 
using nodal surface velocities.  
(c) Semi-Lagrangian advection by mapping from the current configuration to the reference 
configuration, and computing, for each current position, its corresponding 
location in the undeformed configuration and evaluating the reference 
level set field there.  
(d) Interpolation from precomputed level sets parameterized by 
deformation coordinates.  
}
\label{fig:LevelSetUpdateMethods}
\end{figure}

There exist several approaches to update the level set of a particle under deformation~\cite{osher2004level}, i.e., when the surface geometry evolves in time. Each approach comes with its own advantages and limitations, typically involving a trade--off between accuracy and computational expense, both in terms of time and memory complexity. Some methods prioritize strict conservation of the signed distance property, while others emphasize efficiency or ease of implementation. Each of the proposed methods are outlined pictorially in Fig.~\ref{fig:LevelSetUpdateMethods}, and we discuss below their respective benefits, drawbacks, and the contexts in which they are most suitable.

\subsubsection{Recomputation of the level set from the updated particle surface}

The most direct approach is to recompute the level set from the current (deformed) particle 
surface, i.e.,
\begin{equation}
    \phi(\vec{x},t) = 
\begin{cases}
+d(\vec{x}, \Gamma(t)), & \text{if } \vec{x} \text{ is outside the surface}, \\[6pt]
0, & \text{if } \vec{x} \in \Gamma(t), \\[6pt]
-d(\vec{x}, \Gamma(t)), & \text{if } \vec{x} \text{ is inside the surface},
\end{cases}
\end{equation}
where $d(\vec{x},\Gamma(t)) = \min_{\vec{y}\in\Gamma(t)} ||\vec{x}-\vec{y}||$, where $\Gamma(t)$ denotes the particle surface at time $t$. This strategy uses no approximation based on the information about the prior level set or the deformation mode itself. Hence, it is both the most general and the most accurate, since the level set is reconstructed from the exact geometry at every step, and there is no accumulation of potential numerical errors from incremental updates. However, the computational cost is significant, as it requires repeated signed--distance function calculations, and no efficiency gains can be realized from previously computed level sets. This method therefore serves as a natural baseline against which more efficient strategies may be compared.

\subsubsection{Advection-based evolution of the level set}

A widely used strategy in fluid dynamics is to evolve the previously computed level set field by solving a transport (advection) equation driven by the velocity field, in this case associated with the deformation of the particle:
\begin{equation}
\frac{\partial \phi ( \vec{x},t)}{\partial t} + \mathbf{v}( \vec{x},t)\cdot\nabla \phi( \vec{x},t) = 0
\quad\text{in }\Omega,
\label{LsetAdvectionEqn}
\end{equation}
where $\phi( \vec{x},t)$ denotes the level set field, $\mathbf{v}( \vec{x},t)$ is the velocity associated with the particle deformation, which advects the level set in time. The equation is discretized in time (e.g. with an explicit upwind scheme), and solved for $\phi( \vec{x},t^{n+1})$ given the level set $\phi( \vec{x},t^n)$ at the previous time step $t^n$. This framework allows the level set to evolve continuously in time, using information from the prior state rather than recomputing the geometry anew at each step.

A practical issue, on the other hand, in this approach is that the velocity field due to deformation is defined only on the surface and interior of the grain. However, the transport equation is posed over the entire computational grid $\Omega$, since we need to be evolving the level set grid outside the grain boundaries as well, e.g. when computing kinematical measures driving long-range interactions such as bonding forces. This requires the velocity field $\mathbf{u}( \vec{x},t)$ to be specified everywhere, not just on the particle boundary and interior. Since there is no physical basis for velocities outside the material, standard level set practice is to introduce numerical extensions. A common strategy is to extend the velocity field off the interface by extrapolating the known surface velocity into the exterior region, typically along normal directions, so that the transport equation can be consistently evaluated in the required neighborhood of the interface. However, these methods impose additional numerical expense, and their accuracy depends on the choice of extrapolation procedure. Consequently, for applications where accurate modeling of bonding is crucial, this approach presents additional challenges. However, for cases where only contact interactions are of interest, simpler extensions of the level set into narrow bands around the particle surface may be sufficient, thereby reducing both numerical complexity and computational cost. 

It should also be emphasized that this approach remains computationally expensive, since a partial differential equation must be solved at every time step, albeit less costly than the full reinitialization from the current surface geometry. This makes the method suited for simulations of systems with a lower number of particles. While the formulation is general, in the present setting we have access to additional structure: the particle displacements and velocities follow prescribed modal shapes, provided either as closed-form expressions or as numerically extracted fields. This observation suggests that one may construct an alternative update strategy that explicitly harnesses this structure, together with the precomputed initial level set, to achieve improved computational efficiency. This forms the basis of the next method considered.

\subsubsection{Semi-Lagrangian advection of the level set} \label{LagrangianRemapping}

This method is inspired by semi--Lagrangian approaches in fluid mechanics, where particles are traced backward along a velocity field~\cite{osher2004level}. In the present case, the pullback is performed using a known displacement field defined in the local reference frame of the particle.
For each grid node at the current position $ \vec{x}$, we first determine the corresponding 
location in the initial (undeformed) level set field by applying the inverse displacement to the current position, i.e.,
\begin{equation}
 \vec{x}_{\rm ref} =  \vec{x} - \mathbf{u}( \vec{x},t),
\end{equation}
where $\mathbf{u}( \vec{x},t)$ denotes the displacement vector of the material point at time $t$. 
The updated level set is then obtained by interpolating the initial field $\phi_0$ at this mapped location:
\begin{equation}
\phi( \vec{x},t) = \phi_0( \vec{x}_{\rm ref}).
\end{equation}

This mapping is an exact solution of the transport equation~\eqref{LsetAdvectionEqn} for any kinematics of the form $\mathbf{u}( \vec{x},t)=\varepsilon_{\alpha}(t)\,\boldsymbol{\Phi}_{\alpha}( \vec{x})$, up to the discretization of the level set grid. Indeed, by taking the representation $\phi( \vec{x},t)=\phi_0\!\left( \vec{x}-\mathbf{u}( \vec{x},t)\right)$ as an ansatz and substituting it into~\eqref{LsetAdvectionEqn}, one verifies directly---via the chain rule and the identity associated with the inverse deformation map---that it satisfies the advection equation exactly. This approach leverages the known displacement functions together with the initial level set field, thereby providing a computationally efficient strategy while at the same time addressing the challenge of maintaining stability and avoiding the numerical diffusion typically associated with advection schemes.  

However, it is important to note that the same issue encountered in advection-based methods also arises here, namely that the deformation mapping must be applied consistently both inside and outside the grain, even though the deformation displacement field is prescribed only within the interior of the grain. In the case of an analytical closed-form displacement field, the same field can be extended throughout the computational grid. While such an extension preserves numerical consistency, it does not, in general, yield an exact signed-distance representation of the updated geometry outside the grain. In the case of a numerically defined displacement field, the deformation mapping is instead extrapolated into the exterior domain, typically only in the immediate vicinity of the surface~\cite{osher2004level}.  

Both of these strategies are sufficient to maintain stability of the level set evolution and provide a reasonably accurate geometric description near the interface, which is the region relevant for contact computations. For long-range interactions (e.g. bonding forces) that require an accurate level set representation over a larger portion of the exterior domain, however, such extensions become less reliable, and their implications should be examined more carefully. Thus, this limitation is particularly relevant for bonded particle methods. In the present work, we restrict attention to dry-contact scenarios and all numerical examples are conducted without long-range bonding interactions.

Although this semi-Lagrangian approach is more efficient, explicitly deforming and storing the level set over the full grid is still expensive for large grids or many particles. The key observation in LS-DEM is that contact and bonding queries only need the local level set value at a surface node, obtained by trilinear interpolation from the eight surrounding grid nodes of the other grain~\cite{kawamoto2016level}. Since the deformed value at any point depends only on 
the position $ \vec{x}$ and the displacement $\mathbf{u}( \vec{x},t)$, the field never needs a global update. Instead, we remap on demand, namely, for each query, we apply the inverse mapping $\vec{x}_{\rm ref} =  \vec{x}-\mathbf{u}( \vec{x},t)$
 to the eight corner nodes, sample the initial (undeformed) level set at those locations, and use those eight values for the trilinear interpolation. As a result, no global update or solution of a PDE is required over the grid, allowing the deformable LS-DEM to operate within the same order of computational complexity as rigid DEM.

Next, we introduce an alternative methodology that maintains the low computational cost of the present framework, while enabling an accurate representation of the level set field both inside and outside the grain.

\subsubsection{Interpolation along precomputed level sets}

The central idea of this method is to generate, at a preprocessing step, level set fields corresponding to a representative set of deformation states for each deformation mode, based on selected values of the generalized deformation degrees of freedom. During the simulation, the current level set field is then interpolated (e.g. linearly) from these precomputed fields—across all modes. In this approach, the computation reduces to updating only the scalar multipliers (weights) associated with the precomputed fields, rather than solving an advection PDE or performing repeated deformation mappings. 
For a single deformation mode $\alpha$, the contribution of that mode to the
level set field is represented as
\begin{equation}
\phi^{(\alpha)}( \vec{x},t)
=
\sum_{p=1}^{P_i}
w_{\alpha,p}\big(\varepsilon_{\alpha}(t)\big)\,
\phi_{\varepsilon_{\alpha,p}}( \vec{x}),
\label{eq:lset_single_mode}
\end{equation}
where $\phi_{\varepsilon_{\alpha,p}}( \vec{x})$ denotes the precomputed
level set field corresponding to the $p$-th sampled value
$\varepsilon_{\alpha,p}$ of the generalized deformation coordinate
$\varepsilon_{\alpha}$, and $w_{\alpha,p}(\varepsilon_{\alpha}(t))$ is the associated
interpolation weight at time $t$.
 
The total level set field is then obtained by superposition of all deformation modes:
\begin{equation}
\phi( \vec{x},t)
=
\sum_{\alpha=1}^{N_{\text{modes}}}
\sum_{p=1}^{P_i}
w_{\alpha,p}\big(\varepsilon_{\alpha}(t)\big)\,
\phi_{\varepsilon_{\alpha,p}}( \vec{x}).
\label{eq:lset_full_multimode}
\end{equation}

In principle, this approach leads to the lowest on--the--fly computational cost among the proposed methods. As in the semi--Lagrangian formulation, level set values are evaluated only at query points, without performing any grid--wide update of the level set field. More fundamentally, this strategy provides a natural representation of both the interior and exterior level set fields without requiring explicit numerical extrapolations. However, this comes at the cost of an increased memory footprint and a higher preprocessing computational burden.

In the following validation examples (Section~\ref{SectionValidation}), we adopt the semi--Lagrangian advection approach.

\subsection{Particle interactions}
\label{SectionParticleInteractions}

In the proposed deformable LS-DEM framework, particle interactions follow exactly the same contact and bonding formulations as in the rigid LS--DEM method~\cite{feldfogel2024discretization, harmon2021modeling}; the key difference is that both the nodal positions of each grain and the corresponding level set fields evolve in time according to the prescribed deformation modes. Contact forces are computed by evaluating the level set value at each surface node of one grain using its current spatial location within the evolving level set field of the other grain, via trilinear interpolation over the eight grid nodes that surround that location \cite{kawamoto2016level}. Bonding interactions rely on the same node--to--level set evaluation procedure, but require additional kinematic quantities for the constitutive law of the bonding medium \cite{harmon2021modeling}. 

\subsection{Generalized force computation and time integration}
\label{GeneralizedForceComputation}
The first two equations, Eqs.~\eqref{FinalSetOfEquationsDetailed1} and~\eqref{FinalSetOfEquationsDetailed2},
of the governing equations derived in Section~\ref{SectionGoverningEquations}, correspond to the translational and rotational motion, and their implementation follows the standard rigid-body LS--DEM algorithm, with the difference being that the geometric quantities entering the rotational balance, such as the lever arms and local surface directions, depend on the current deformation state. Eq.~\eqref{FinalSetOfEquationsDetailed3} governs the evolution of the deformation degrees of freedom, and involves the calculation of generalized forces, by evaluating the dot product between the force acting at a node and the unit vector in the direction of the displacement field. Given that particle deformability is described here by fixed mode shapes defined in the undeformed local configuration of each grain, the corresponding displacement vectors are obtained by evaluating the respective shape functions using the undeformed coordinates of the associated surface node. Note, however, that interparticle forces are computed in the global frame, whereas the modal displacement directions are defined in the local frame of the grain. Hence, care must be taken to first map the force vectors from the global frame to the local frame of the grain, accounting for rigid-body rotation, before computing the dot product.
Finally, once the generalized forces have been assembled, the equations of motion for the deformation degrees of freedom are advanced in time using an explicit integration scheme. The critical time step is chosen to satisfy the most stringent of two critical time steps, $\Delta t_{\text{rigid}} = \beta_r \sqrt{k_n/m_{\text{eff}}}$ for the rigid motion of the particle (where $k_n$ is the particle interaction stiffness and $m_{\text{eff}}$ is the effective mass of a pair of particles), and $\Delta t_{\text{deform}} = \beta_d/\omega_{\text{max}}$ (where $\omega_{\text{max}}$ is the largest eigenfrequency of the dynamical system governing the generalized deformation degrees of freedom (Eq.~\ref{linearTensorialEquations3})). The coefficients $\beta_r$ and $\beta_d$ are chosen empirically as conservative safety factors. Finally, a modal damping term may be included to enhance numerical stability and to suppress spurious high-frequency oscillations.

\section{Examples and Validation}
\label{SectionValidation}

This section illustrates the application of the proposed deformable LS--DEM framework and its validation through a series of representative examples. The aim is to demonstrate the consistency of the formulation across different deformation modes, either prescribed analytically through closed-form expressions (Section~\ref{analyticalModesSection}) or, more generally, obtained from finite element simulations (Section~\ref{NumericallyDerivedModesSection}). 

\subsection{Analytical modes}
\label{analyticalModesSection}

Analytical modes are suitable when grains are expected to exhibit a deformation mode represented by an \textit{a priori} closed-form mathematical expression. This is true, for instance, when modeling particulate systems composed of slender topologically interlocked building blocks—such as biological structural materials like nacre (Fig.~\ref{fig:ParticulateMatter}). Stretching and bending modes would be expected to represent a dominant mode of deformation of these grains. 

\subsubsection*{Bending of slender grains} 
\label{AnalyticalBending}
\begin{figure}[t!]
    \centering
    \begin{minipage}{1.0\linewidth}
        \centering
        \includegraphics[width=\linewidth,
                         trim={0cm 0cm 0cm 0cm},clip]
                         {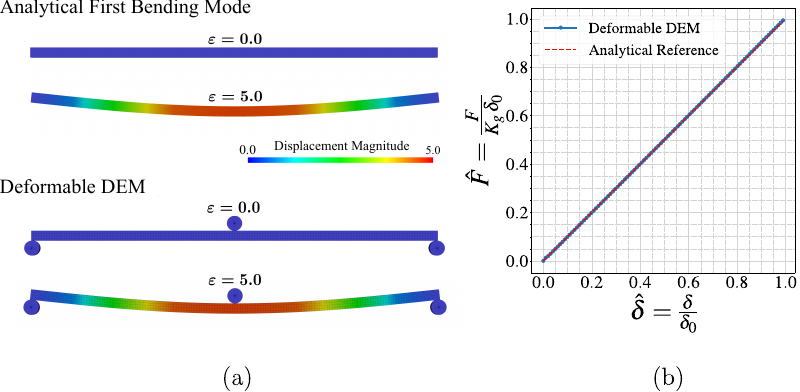}
    \end{minipage}
\caption{Three-point bending benchmark test for an analytical deformation mode. 
(a) Undeformed and deformed configuration of the analytical first bending mode, shown together with the grain obtained from the deformable LS-DEM simulation. The analytical mode is scaled to the same modal amplitude for qualitative comparison of the deformation pattern, as defined by the displacement field in Eq.~\ref{eq:bending_mode_shape}. Color indicates the magnitude of the displacement field. (b) Quantitative comparison between the numerical normalized force–displacement response and the analytical stiffness reference derived from the generalized stiffness.}
\label{fig:ThreePointBending}
\end{figure}

In this first example, we consider the bending of a slender deformable cuboidal particle.
The particle is simply supported on two rigid cylinders at its ends, and subjected to three-point bending as a result of indentation by a third rigid cylinder (indenter) at the top, as shown in Fig.~\ref{fig:ThreePointBending} a). In this simple problem, the bending mode is described by a closed-form analytical expression for the displacement field, given by
\begin{equation}
\begin{aligned}
u_x &= z\,\varepsilon\,\frac{\pi}{L}\,\cos\!\left(\frac{\pi x}{L}\right), \\
u_z &= -\,\varepsilon\,\sin\!\left(\frac{\pi x}{L}\right), \\
u_y &= 0,
\end{aligned}
\label{eq:bending_mode_shape}
\end{equation}
where $\varepsilon$ denotes the modal amplitude (the value of the generalized displacement) and $L$ the grain length in the bending plane. Importantly, the coordinates $(x,y,z)$ denote the undeformed coordinates of a material point of the grain, in accordance with the reduced kinematic description adopted in the present framework, in which fixed shape functions are defined over the reference configuration. Here, we adopt the first bending mode corresponding to the fundamental vibration mode of a simply supported Euler--Bernoulli beam for its mathematical simplicity. Alternatively, one could use the quasi-static Euler--Bernoulli solution with free-end boundary conditions, which yields nearly identical generalized stiffness, mass, and deformation shapes. The expressions of Eq.~\ref{eq:bending_mode_shape} are used both to deform the surface mesh nodes of the grains and the level set field --- following the semi-Lagrangian advection approach of Section~\ref{LagrangianRemapping} --- and to compute the generalized stiffness, corresponding in this case to a quadratic elastic potential (Eq.~\ref{ModalStiffnessDefinition}), and mass (Eq.~\ref{ModalMassDefinition}).

These closed-form expressions are also used to obtain the deformation displacement vectors required for the computation of generalized forces (Eq.~\ref{FinalSetOfEquationsDetailed3}) arising due to contact.

Towards a quantitative validation, the displacement $\delta$ and reaction force $F$ at the indenter are computed throughout the simulation. The numerical parameters used in this validation example are summarized in \ref{Appendix:ValidationParameters}, Table~\ref{table:InputsForBendingExample}. Fig.~\ref{fig:ThreePointBending} b) compares the force--displacement response measured at the indenter with the analytical generalized stiffness associated with the bending mode~($K_g$), showing excellent agreement. The displacement field of the mode shape is normalized such that the generalized displacement amplitude corresponds to the vertical displacement at the mid-span of the beam, which coincides with the displacement of the indenter. Accordingly, the displacement field given in Eq.~\ref{eq:bending_mode_shape} is used directly, with \(u_z = \varepsilon\) at \(x = L/2\) and \(u_x = u_y = 0\) at \(x = L/2\). 
Both force and displacement are expressed in nondimensional form as
\begin{equation}
\hat{\delta} = \frac{\delta}{\delta_0}, \ \ \
\hat{F} = \frac{F}{K_g\,\delta_0},
\label{normalizationForPlotting}
\end{equation}
where \(\delta_0\) is the reference displacement corresponding to the maximum indentation applied at the indenter, and \(K_g\) denotes the generalized stiffness of the bending mode. Under this normalization, the analytical prediction reduces to \(\hat{F} = \hat{\delta}\); that is, the slope of the normalized force--displacement response measured at the indenter should be unity when the normalization is applied. 

Overall, this validation exercise establishes that the generalized degree of freedom corresponding to an analytically defined deformation mode evolves in time as intended under the action of the generalized forces driven by contact interactions with neighboring particles, and that it manifests as correctly displaced surface nodes and a correctly updated level set.

\subsection{Numerical modes}
\label{NumericallyDerivedModesSection}

Analytical deformation modes, while valuable when available, are often not known \emph{a priori} for particulate systems subjected to complex boundary and loading conditions. In many practical settings, grain-scale deformation mechanisms cannot be inferred from physical intuition alone and do not admit closed-form expressions. This motivates the introduction of numerically derived deformation modes, which are extracted from auxiliary high-fidelity simulations—hereafter referred to as \emph{mode-extraction simulations}—or experiments, and subsequently incorporated into the deformable DEM framework. The extraction of deformation modes can be approached at different representative scales depending on the nature of the loading, boundary, and configurational conditions expected to trigger \emph{bulk grain deformation}, as well as on the degree of prior knowledge of these physical scenarios.
\begin{itemize}

    \item \textit{Extraction from a unit cell}\\
    When the granular system exhibits known periodicity or structural order, a repeating unit cell can be identified, which is composed of a small number of particles. High-fidelity FEM simulations performed on this unit cell under representative loading conditions allow direct extraction of the dominant particle deformation modes. This approach is well suited to granular crystals and periodic nacre-like platelet assemblies, where deformation mechanisms repeat throughout the system.

    \item \textit{Extraction from a representative volume element}\\
    In systems lacking clear periodicity or where deformation is governed by collective mesoscale effects, deformation modes can be identified from FEM simulations on statistically representative volume elements (RVEs). These are chosen large enough to capture the relevant physics - such as morphology, density, contact topology, and material heterogeneity - yet smaller than the full system. By sampling multiple such volumes and monitoring the convergence of extracted modes with respect to domain size and ensemble size, a statistically representative basis of deformation modes may be constructed. 

    \item \textit{Extraction informed by system-wide behavior}\\
    In certain systems, bulk grain deformation is not governed by local or mesoscale mechanisms, but instead emerges only under global features such as boundary conditions (e.g. jamming percolation). When the influence of particle deformation is limited, auxiliary rigid DEM simulations may be used to identify the interparticle forces driving grain deformation, which can in turn be used to obtain the dominant deformation modes. Conversely, when this effect is pronounced, the deformation should be extracted directly from a fully resolved simulation of the complete system. Although the cost is high, these modes, once extracted, may be used efficiently within the deformable DEM framework, e.g., for parametric investigation.

     \item \textit{Extraction from experiments}\\ 
     Experimental approaches provide an additional route for mode extraction, allowing deformation modes to be obtained directly from measured displacement fields. Techniques such as X-ray computed tomography combined with digital image correlation can provide three-dimensional deformation data, from which dominant modes can be extracted analogously to the aforementioned numerical approaches.
     
\end{itemize}

Any of the above strategies produce a set of displacement fields \{$\boldsymbol{u}_i\}_{i=1}^{N}$ that may represent snapshots from one mode-extraction simulation under increasing levels of deformation, or snapshots from multiple simulations. Beyond the displacement fields, these data also provide measurements of the associated strain energies (in the case of simulations) and reaction forces (in the case of simulations or experiments). Based on those measurements, the corresponding modes and associated generalized forces are computed, which depend on the nature of the problem at hand as discussed in detail below. In any case, the set of modes should be sufficiently rich to capture the dominant deformation mechanisms, and linearly independent in order to avoid numerical degeneracies in the reduced system. Moreover, depending on the importance of inertial effects on the deformation response, the deformation modes should be extracted from quasi-static simulations or time-resolved dynamic simulations/modal analyses. In the following, we focus on the extraction of quasi-static modes.

\subsubsection*{Linear problems}

When the mode-extraction problem is linear, the displacement field inside a particle scales linearly with the applied loading, leading to a linear generalized force--displacement relationship and a constant generalized stiffness. When only a single mode is considered, it is directly obtained as the displacement field extracted at any representative equilibrium configuration. The subsequent use of this deformation shape function is analogous to the approach on analytical modes, with the difference being that now the shape function is defined on a mesh and, hence, any displacement values (e.g. for displaced surface nodes, level set update, evaluation of generalized force) are obtained by interpolation. The generalized stiffness may be obtained directly from the simulation as the slope of the force--displacement response at the load application point, or by evaluating Eq.~\ref{ModalStiffnessDefinition} as a numerical integral using the numerically defined shape function. 

When multiple linear problems are considered, then the corresponding displacement fields cannot, in general, be directly used as modes, as they are not guaranteed to be linearly independent or to form a suitable, well-conditioned reduced basis; instead, a basis of modes $\{\boldsymbol{\Phi}_\alpha(\boldsymbol{x})\}_{\alpha=1}^{A}$ (where $A$ is the number of considered modes) must be constructed. The available displacement fields \{$\boldsymbol{u}_i\}_{i=1}^{N}$ that represent $N$ snapshots from mode-extraction simulations are collected column-wise into a matrix $\boldsymbol{U}$, whose Singular Value Decomposition (alternatively Proper Orthogonal Decomposition) is subsequently computed. The chosen $A$ left singular vectors then furnish the desired modes $\{\boldsymbol{\Phi}_\alpha(\boldsymbol{x})\}_{\alpha=1}^{A}$. Next, the generalized internal elastic force associated with each mode $\boldsymbol{\Phi}_\alpha(\boldsymbol{x})$ must be determined. The total reduced strain energy for a particle takes a quadratic form
\begin{equation}
    U_{\varepsilon} = \frac{1}{2} \varepsilon_\alpha K_{\alpha\beta} \varepsilon_\beta,
    \label{}
\end{equation}
where the reduced stiffness is obtained e.g. by energy integration
\begin{equation}
K_{\alpha\beta} = \int_{\Omega} \bigl(\mathbb{C}:\boldsymbol{E}(\boldsymbol{\Phi}_{\alpha})\bigr) :
\boldsymbol{E}(\boldsymbol{\Phi}_{\beta})\,\mathrm{d}V,
\label{}
\end{equation}
and where $\Omega$ denotes the grain’s reference configuration, and $\boldsymbol{E}(\boldsymbol{\Phi}_{\alpha})$ is the strain field induced by the $\alpha$-th deformation mode. 
Accordingly, the internal generalized force associated with mode $\alpha$ is given by
\begin{equation}
F_{\varepsilon_\alpha}^{\text{int}} = \frac{\partial U_\varepsilon}{\partial \varepsilon_\alpha} = K_{\alpha\beta} \varepsilon_\beta
\end{equation}
Note that the deformation modes obtained through the above procedure are, by construction, orthogonal in a vectorial (Euclidean) sense. This orthogonality guarantees a compact and well-conditioned reduced basis, but does not, in general, imply orthogonality with respect to the elastic energy. If desired, the modes may also be orthogonalized in an energetic sense, leading to an additive decomposition of the deformation energy and a diagonal generalized stiffness, and hence to a set of decoupled modal evolution equations. The same is true when modes are furnished by a generalized eigendecomposition of an FEM model~\cite{williams1987modal}, though in that case the modes are particle-intrinsic rather than tailored to specific deformation mechanisms, and, hence, the resulting basis may be less efficient than the proposed construction.

\subsubsection*{Extension to nonlinear problems}

In the case of nonlinear mode extraction problems (e.g., due to material elastic nonlinearity or pronounced evolving interparticle contact), the associated displacement field evolves nonlinearly with the applied loading, accompanied by a nonlinear relationship between the generalized force and the generalized displacement, and a nonquadratic underlying elastic potential energy. In the present work, consistent with the small-deformation assumptions adopted in the theoretical framework, we focus on nonlinearities induced by evolving contact, while particle behavior is assumed to remain linear elastic and governed by small-strain kinematics. These effects can be accommodated within the current formulation through a nonlinear elastic potential energy description. By contrast, incorporating material nonlinearity associated with finite particle deformation would require relaxing the small-strain assumption and accounting for deformation-dependent changes in the generalized mass and inertia that naturally arise under such large shape changes. These are discussed as future extensions of the framework.

The general strategy for the construction of the modal basis $\{\boldsymbol{\Phi}_\alpha(\boldsymbol{x})\}_{\alpha=1}^{A}$ is identical to the one discussed in the case of linear problems. The difference  is that the displacement fields used to construct the modes represent simulation snapshots under increasing levels of deformation given the nonlinearity of the problem, and as a result more modes might, in general, be required to achieve an accurate description of the particle deformation. However, in the case where loading is represented by one dominant path, then, instead of this general strategy, an efficient basis may be constructed by a single mode, taken as the displacement field at a large representative deformation, i.e. $\boldsymbol{\Phi}_{\alpha}(\boldsymbol{x}) = \boldsymbol{u}(\boldsymbol{x},\varepsilon^{\max})$. This procedure is in fact illustrated in the subsequent validation example. In case this single mode proves inadequate, it may be supplemented by modes representing the differences between subsequent deformation states, i.e. $\boldsymbol{\Phi}_\alpha(\boldsymbol{x}) = \boldsymbol{u}(\boldsymbol{x},\varepsilon_{\alpha}) - \boldsymbol{u}(\boldsymbol{x},\varepsilon_{\alpha-1})$, termed \emph{difference modes}, and subsequently pruned/orthogonalized as needed. The choice of basis can be tested by additional representative out-of-sample validation simulations, and measuring the error in terms of displacement or strain energy reconstruction.

On the other hand, the construction of the energy and generalized forces for nonlinear problems is more complex. When total strain energy measurements $PE_e^i$ are available in each snapshot $i$ (e.g. in an FEM calculation), and for each value of the corresponding generalized deformation coordinates $\varepsilon_\alpha^i$, then a parametric energy $U_\varepsilon^\theta(\varepsilon_\alpha)$ in terms of parameters $\theta$ (the simplest option being a tabulation) may be constructed, giving rise to a least squares problem
\begin{equation}
    \min_{\theta} \sum_i (U_\varepsilon^\theta(\varepsilon_\alpha^i) - PE_e^i)^2.
    \label{}
\end{equation}
The associated internal generalized force for mode $\alpha$ follows as
\begin{equation}
    F_{\varepsilon_\alpha}^{\text{int}} = \frac{\partial U_\varepsilon^\theta}{\partial \varepsilon_\alpha}.
    \label{}
\end{equation}
In the absence of energy measurements (e.g. in experimental mode extraction), we can compute the generalized forces from measured reaction forces, by fitting those to the gradient of the elastic potential, namely
\begin{equation}
    \min_{\theta} \sum_i (\boldsymbol{\nabla}_{\boldsymbol{\varepsilon}} U_\varepsilon^\theta(\varepsilon_\alpha^i) - \boldsymbol{F}_{\text{react}}^i)^2.
    \label{}
\end{equation}
In both cases, by expressing the strain energy as a state function of the generalized deformation coordinates, this approach guarantees conservativeness by construction. Crucially, it can capture not only material elastic nonlinearities but also nonlinearities arising due to evolving contact constraints. Of course, in the case of one generalized coordinate $\varepsilon$, the associated generalized force $F_\varepsilon$ is obtained directly as the reaction force.

\subsubsection*{Compaction of spherical grains}
\label{ref:CompactionOfSphericalGrains}

\begin{figure}
    \centering
    \begin{minipage}{1.0\linewidth}
        \centering
        \includegraphics[width=\linewidth,
                         trim={0cm 0cm 0cm 0cm},clip]
                         {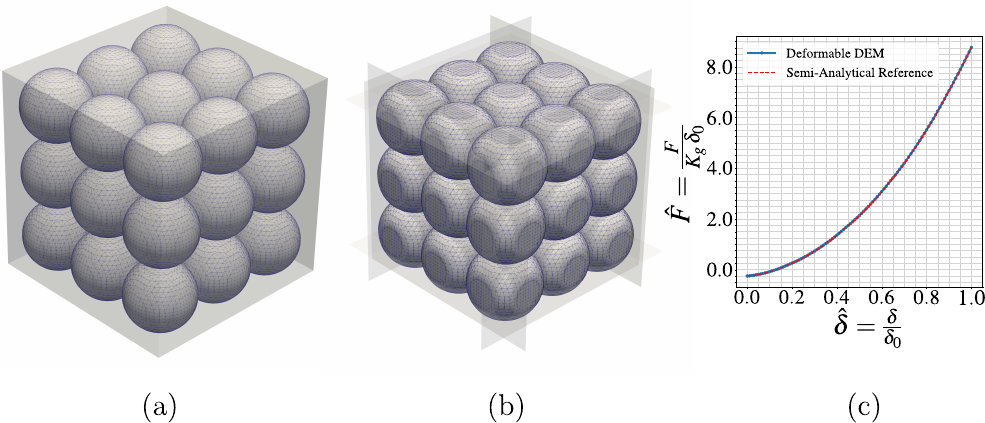}
    \end{minipage}

    \caption{System-level compaction of spherical grains using the numerically extracted deformation mode.  
    (a) Undeformed configuration of the 27-sphere assembly.  
    (b) Deformed and compacted configuration of the 27-sphere assembly under hydrostatic compression.
    (c) Quantitative comparison between the deformable DEM force–displacement response and the semi-analytical reference.} 

    \label{fig:SetOfCompactedSpheres}
\end{figure}

As an illustrative case of numerically derived nonlinear modes, we consider the example of a compaction of an ordered packing of frictionless spherical grains under hydrostatic compression, as shown in Fig.~\ref{fig:SetOfCompactedSpheres} a). This example highlights the ability of the deformable LS-DEM formulation to capture grain-scale volumetric deformation and contact flattening, leading to densification, which is central to granular mechanics but inaccessible to conventional rigid-particle DEM. 

\begin{figure}[t!]
    \centering
    \begin{minipage}{1.0\linewidth}
        \centering
        \includegraphics[width=\linewidth,
                         trim={0cm 0cm 0cm 0cm},clip]
                         {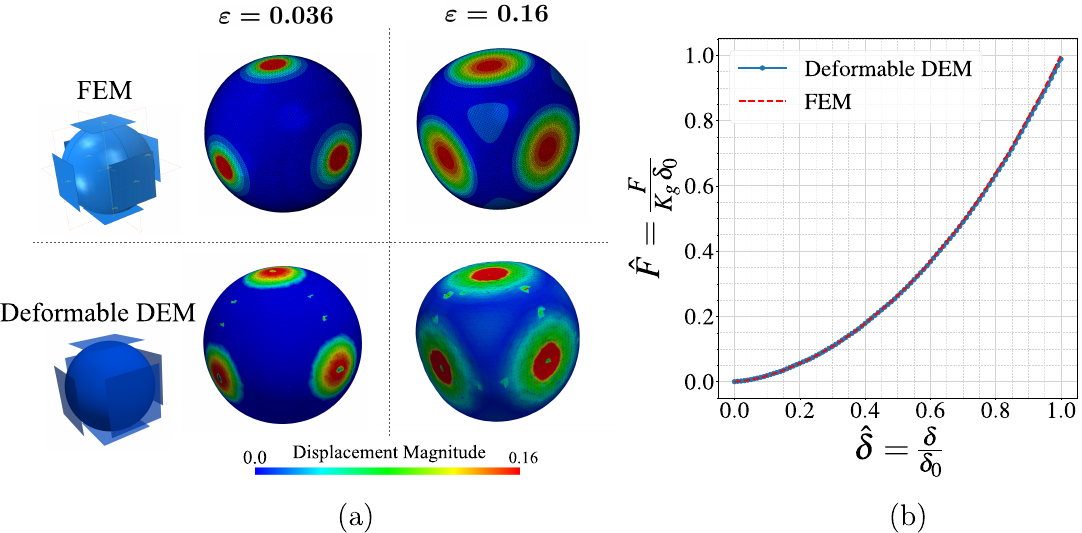}
    \end{minipage}

\caption{Comparison between the results obtained from the high-fidelity FEM indentation simulation and the results obtained from the deformable LS-DEM framework under the same indentation setup, using a deformation mode extracted from the FEM simulation. 
(a) Deformed shapes obtained in FEM and in deformable LS-DEM for the same value of the generalized modal displacement used to scale the extracted deformation mode. Color indicates the magnitude of the displacement field; the same color scale is used for both representations. 
(b) Quantitative comparison between the normalized force–displacement response from the deformable LS-DEM simulation and the FEM force–displacement curve.
}
   \label{fig:DEM_FEM_Comparison}
\end{figure}

Following the proposed unit cell-based mode-extraction strategy for this periodic system, the dominant grain deformation is obtained from a fully-resolved FEM simulation of a deforming sphere. The periodicity of the problem admits a description of the kinematics by a displacement field symmetric along three principal orthogonal directions.  Due to symmetry, sphere-to-sphere contact may be equivalently represented by sphere-to-rigid flat surface contact. Accordingly, we consider the indentation of a linear elastic sphere by six orthogonal frictional rigid walls (indenters),  carried out in \textsc{Abaqus}, as illustrated in Fig.~\ref{fig:DEM_FEM_Comparison}.

Although the material response in this mode-extraction simulation is linear elastic and governed by small-strain kinematics, the contact conditions induce strong nonlinearity: as indentation proceeds, the contact radius expands, and the displacement field evolves nonlinearly with loading. In principle, the evolving displacement field during compaction is best approximated in our reduced-order framework by a superposition of deformation modes (e.g. difference modes), as described in Section~\ref{NumericallyDerivedModesSection}. Here, however, we show that accurate results can still be obtained even with a single deformation mode extracted from the FEM displacement field, provided it is chosen at an appropriate level of deformation to guarantee physical consistency. In this case, the mode is extracted from the displacement field at maximum indentation/compaction, which preserves the characteristic contact-flattening pattern when scaled. Crucially, the particle deformation determines the elastic energy stored in the sphere, so that the resulting Hertzian force–displacement response emerges from the shape change rather than being imposed as an effective contact law.

In general, to determine the generalized force for the system, one needs to evaluate the energy $U_\varepsilon(\varepsilon)$, and its derivative at the current level of generalized deformation $\varepsilon$, as described in the previous Section. Alternatively, here, for this frictionless problem and given that the indenter displacement coincides with the generalized coordinate $\varepsilon$, the measured reaction force directly corresponds to the generalized force $F_\varepsilon = U'_E(\varepsilon)$, while the generalized stiffness coincides with the local tangent stiffness of the corresponding force-displacement response at a given level of deformation. The reaction forces are computed during the simulation and stored to a look-up table (see Appendix~B) to be later used in the deformable LS-DEM simulation, when integrating the evolution of $\varepsilon$ (Eq.~\ref{FinalSetOfEquationsAbstract3}). This avoids explicitly reconstructing the elastic potential $U_\varepsilon(\varepsilon)$.

Using the identified mode shape and corresponding generalized force, a single-particle compression, identical to that employed in the fully resolved FEM simulation, is carried out in the deformable LS-DEM. Fig.~\ref{fig:DEM_FEM_Comparison} compares the deformation shapes obtained from the FEM simulation and from the deformable LS-DEM, as well as the corresponding force--displacement responses. A normalization procedure analogous to that adopted for the analytical bending case is employed (see Eq.~\ref{normalizationForPlotting}); here, the generalized stiffness \(K_g\) is obtained by evaluating the modal stiffness definition in Eq.~\ref{ModalStiffnessDefinition} using the mode extracted at the final loading step, which coincides with the local tangent stiffness of the corresponding force--displacement response at that step. The evolution of the contact region and the associated surface flattening are captured accurately by the reduced model. Minor deviations are observed at very small generalized displacements as a direct consequence of the fixed-shape approximation. The excellent agreement observed in both the deformation geometry and the reaction forces confirms the correct embedding and activation of the numerically derived compaction mode within the deformable LS-DEM formulation. The simulation is repeated exchanging master and slave grain roles for the deformable particle and the rigid flat indenters, exhibiting overall the same behavior. The numerical parameters used in this example are summarized in \ref{Appendix:ValidationParameters}, Table~\ref{table:InputsForNumericalMode}.

Next, using the same numerically extracted compaction mode within the deformable DEM framework, we perform a system-level simulation to illustrate the response of a granular assembly subject to isotropic compression, as shown in Fig.~\ref{fig:SetOfCompactedSpheres} a) and b). We consider a simple configuration consisting of 27 spherical grains arranged in a cubic packing as described previously, which is compressed from all six orthogonal directions by rigid walls. In this case, each rigid wall is displaced by an amount corresponding to three times the maximum generalized modal amplitude used in the single-particle tests, i.e.\ \(u_w=3\times\varepsilon\), consistent with the kinematic relation derived below.  

To quantitatively validate the force--displacement response obtained from the deformable DEM simulation of the compacted sphere assembly, we examine the energetic consistency of the system and derive a semi-analytical expression for the total force exerted by the six surrounding rigid walls on the particle pack as a function of the imposed wall displacement. Owing to the symmetry of the configuration and the prescribed kinematic constraints, this relation can be obtained directly from the single-particle compaction response without recourse to a system-level FEM simulation. The resulting relation is derived in detail in Appendix C, and reads
\begin{equation}
F_w(u_w) = 9\,F^{\mathrm{int}}_{\varepsilon}\!\left(\tfrac{u_w}{3}\right),
\end{equation}
where \(F_w\) denotes the total force acting on the rigid walls and \(F^{\mathrm{int}}_{\varepsilon}\) is the internal generalized force associated with the compaction mode, evaluated as a function of the wall displacement through the kinematic relation \(\varepsilon = u_w/3\). We adopt the same normalization as in the single-particle compaction case. The deformable DEM response exhibits excellent agreement with this semi-analytical prediction, as shown in Fig.~\ref{fig:SetOfCompactedSpheres} c).

\section{Discussion and Conclusion}\label{SectionDiscussion}
        
We have developed a variational framework for explicit particle-scale simulations of granular matter that accounts for particle deformation through a reduced-order description. Building on fundamental variational principles, the proposed formulation extends rigid-body dynamics to incorporate deformability, while retaining the favorable computational structure and efficiency of classical rigid DEM. In contrast to existing deformable DEM approaches, the proposed framework is based on a modal decomposition that avoids stringent kinematic restrictions and accommodates general deformation shapes defined on particles of arbitrary geometry and topology, while admitting nonlinear elastic effects. These features stem from the generality of the theoretical formulation and its efficient level set-based implementation. Finally, different strategies for level set evolution and mode extraction are proposed, each with its own advantages and disadvantages.

The proposed deformable LS-DEM framework is not without limitations. First, it is not particularly well suited for extreme particle deformations, where fully resolved simulations may be warranted instead. Although any deformation is in principle attainable with the reduced-order description, in practice, the required number of modes may grow to a degree that the computational advantage of this framework is lost. Second, the identification of relevant deformation modes may require costly analyses or physical insight, while care must be taken to avoid under-complete or mutually dependent modes. Insufficient or highly correlated modes may lead to biased energy partitioning or numerical degeneracy in the reduced dynamics. Third, for nearly isotropic grains (e.g., spheres), several deformation modes are rotationally degenerate, meaning the same modal shape can appear in any orientation with the same energy. With a fixed modal basis defined in the local frame of the grain, the model currently requires a correct alignment between the interparticle forces and the chosen mode orientations in order to activate that mode.

It is finally worth noting the multitude of possible avenues for extending the method. To begin with, although the present formulation is derived under a small-strain elasticity assumption, the generality of the underlying variational structure provides clear and systematic pathways for extension to large-deformation settings. The simplest such extension consists in relaxing the small-strain assumption while retaining a reduced kinematic description with fixed shape functions. This entails retaining the deformation-dependent derivative terms of the generalized mass and moment of inertia, leading to extended governing equations with additional inertial contributions. This extension is useful as long as the dominant deformation mechanisms are well represented with a reasonable number of fixed shape functions. Another interesting extension concerns the treatment of the aforementioned rotational degeneracy under particle symmetries. Instead of the simple yet inefficient solution of enriching the modal representation by multiple rotated versions of the same mode, we envision a symmetry-aware extension in which deformation subspaces are endowed with additional internal rotational degrees of freedom. Finally, the present energetic variational structure also provides a basis for readily incorporating additional physical mechanisms, including coupled multiphysics and inelasticity. This would require the incorporation of additional reduced-order generalized dissipative variables in close analogy with similar developments in continuum mechanics, with limited changes to the fundamental structure of the current mathematical derivation.

Overall, the proposed deformable LS-DEM framework holds potential for efficiently modeling the mechanics of a wide range of natural and engineered systems, offering significant acceleration compared to fully resolved techniques. These systems range from periodic architected packings and granular metamaterials, where particle-level deformation may govern macroscopic behavior, to disordered soft matter and jammed assemblies, where heterogeneous contact networks give rise to rich emergent behavior. In both regimes, the developed framework enables efficient parametric studies, while facilitating data-informed exploration and design of particulate systems.

\section*{Acknowledgements}
The authors acknowledge fruitful discussions with Prof. Shai Feldfogel, during earlier conceptions of this computational framework.

\section*{Author contributions: CRediT}
Thomas Henzel: Conceptualization, Formal analysis, Investigation,  Methodology, Software, Validation, Visualization, Writing -- original draft.

Konstantinos Karapiperis: Conceptualization,  Methodology, Supervision, Writing – review \& editing.   

\section*{Declaration of generative AI and AI-assisted technologies in the manuscript preparation process}
During the preparation of this work the authors used ChatGPT (OpenAI) to improve language clarity and readability. After using this tool, the authors reviewed and edited the content as needed and take full responsibility for the content of the published article.

\section*{Declaration of competing interests}
The authors declare that they have no known competing financial interests or personal relationships that could have appeared to influence the work reported in this paper.

\section*{Funding}
This research did not receive any specific grant from funding agencies in the public, commercial, or not-for-profit sectors.

\section*{Data availability}
The data and code supporting the findings of this study are available from the corresponding author upon reasonable request.

\appendix

\section{Variational derivation of the governing equations}
\label{appendixVariationalDerivation}

This appendix provides the detailed derivation connecting the variational expression in Eq.~\ref{StationaryConditionEpsilonDerivative} to the weak form reported in Eq.~\ref{ExpressionDerivedInAppendix}. We start from the variation of the action
\begin{gather}\label{StationaryConditionAppendix}
\hspace{-1.0cm}\delta S =
\frac{d}{d\epsilon} \Bigg[
\int_{t_1}^{t_2} \Big(
\tfrac{1}{2} M_i (\dot{X}_i + \dot{\epsilon \eta_{x_i}} )^2
+ \tfrac{1}{2} I_i(\varepsilon_{\gamma} + \epsilon \eta_{\varepsilon_{\gamma}})
  (\dot{\Theta}_i + \dot{\epsilon \eta_{\Theta_i}} )^2
 \nonumber\\[4pt]
\hspace{-1.25cm}\qquad\qquad
+ \tfrac{1}{2} M_{\alpha\beta}(\varepsilon_{\gamma}\!+\!\epsilon \eta_{\varepsilon_{\gamma}})
  (\dot{\varepsilon}_{\alpha}\!+\!\dot{\epsilon \eta_{\varepsilon_{\alpha}}})
  (\dot{\varepsilon}_{\beta}\!+\!\dot{\epsilon \eta_{\varepsilon_{\beta}}})
- U_\varepsilon(\varepsilon_{\alpha}+\epsilon \eta_{\varepsilon_{\alpha}})
\Big) \,\mathrm{d}t \Bigg]_{\epsilon = 0}.
\end{gather}

Since the variation parameter $\epsilon$ and time $t$ are independent variables, differentiation with respect to $\epsilon$ commutes with time integration. Thus, we can bring the derivative inside the integral to get
\begin{gather}\label{StationaryConditionEpsilon0}
\hspace{-0.3cm}\delta S =
\Bigg[
\int_{t_1}^{t_2} \Big(
M_i (\dot{X}_i + \dot{\epsilon \eta_{x_i}})\,\dot{\eta}_{x_i}
+ \tfrac{1}{2} I_i'(\varepsilon_{\gamma} + \epsilon \eta_{\varepsilon_{\gamma}})
  \eta_{\varepsilon_{\gamma}} (\dot{\Theta}_i + \dot{\epsilon \eta_{\Theta_i}})^2
  \nonumber\\[4pt]
\qquad\qquad
\hspace{-1.3cm}+I_i(\varepsilon_{\gamma}\!+\!\epsilon \eta_{\varepsilon_{\gamma}})
(\dot{\Theta}_i\!+\!\dot{\epsilon \eta_{\Theta_i}})\,\dot{\eta}_{\Theta_i}
+ \tfrac{1}{2} M_{\alpha\beta}'(\varepsilon_{\gamma}\!+\!\epsilon \eta_{\varepsilon_{\gamma}})
\eta_{\varepsilon_{\gamma}} (\dot{\varepsilon}_{\alpha}\!+\!\dot{\epsilon \eta_{\varepsilon_{\alpha}}})
(\dot{\varepsilon}_{\beta}\!+\!\dot{\epsilon \eta_{\varepsilon_{\beta}}})
\nonumber\\[8pt]
\qquad\qquad
\hspace{-1.0cm}+ \tfrac{1}{2} M_{\alpha\beta}(\varepsilon_{\gamma} + \epsilon \eta_{\varepsilon_{\gamma}})
  \dot{\eta}_{\varepsilon_{\alpha}} (\dot{\varepsilon}_{\beta} + \dot{\epsilon \eta_{\varepsilon_{\beta}}})
+ \tfrac{1}{2} M_{\alpha\beta}(\varepsilon_{\gamma} + \epsilon \eta_{\varepsilon_{\gamma}})
  (\dot{\varepsilon}_{\alpha} + \dot{\epsilon \eta_{\varepsilon_{\alpha}}})\,\dot{\eta}_{\varepsilon_{\beta}}
\nonumber\\[4pt]
\qquad\qquad
- U_\varepsilon'(\varepsilon_{\alpha} + \epsilon \eta_{\varepsilon_{\alpha}})\,\eta_{\varepsilon_{\alpha}}
\Big) \,\mathrm{d}t
\Bigg]_{\epsilon=0},
\end{gather}
where the prime notation in $I_i'(\varepsilon_{\gamma} + \epsilon \eta_{\varepsilon_{\gamma}})$ is used to denote the derivative of the function with respect to its argument, evaluated in the direction of the admissible variation. Similarly, this notation is adopted for the functions $M_{\alpha\beta}$ and $U_\varepsilon$. 

Enforcing $\epsilon $ = 0 yields
\begin{gather}
\delta S =
\int_{t_1}^{t_2} \Big(
M_i \dot{X}_i\, \dot{\eta}_{x_i}
+ \tfrac{1}{2} I_i'(\varepsilon_{\gamma})\, \eta_{\varepsilon_{\gamma}}\, \dot{\Theta}_i^{\,2}
+ I_i(\varepsilon_{\gamma})\, \dot{\Theta}_i\, \dot{\eta}_{\Theta_i}
+ \tfrac{1}{2} M_{\alpha\beta}'(\varepsilon_{\gamma})\, \eta_{\varepsilon_{\gamma}}\, \dot{\varepsilon}_{\alpha} \dot{\varepsilon}_{\beta}
\nonumber\\
\qquad\qquad
+  M_{\alpha\beta}(\varepsilon_{\gamma})\, \dot{\eta}_{\varepsilon_{\alpha}} \dot{\varepsilon}_{\beta}
- U_\varepsilon'(\varepsilon_{\alpha})\,\eta_{\varepsilon_{\alpha}}
\Big) \,\mathrm{d}t,
\label{StationaryConditionVersionCurrentAppendix}
\end{gather}
where we have used the symmetry of the generalized mass matrix to combine the terms with $\frac{1}{2}M_{\alpha\beta}(\varepsilon_{\gamma})\dot{\varepsilon_{\alpha}}\dot{\eta}_{\varepsilon_{\beta}}$ into a single contribution. Further applying integration by parts
\begin{equation}
\int_{t_1}^{t_2} M_i\, \dot{X}_i\, \dot{\eta}_{x_i} \,\mathrm{d}t
= M_i \Big[ \dot{X}_i\, \eta_{x_i} \Big]_{t_1}^{t_2}
- \int_{t_1}^{t_2} M_i\, \ddot{X}_i\, \eta_{x_i} \,\mathrm{d}t,
\label{StationaryCondition1}
\end{equation}
\begin{gather}
\int_{t_1}^{t_2} I_i(\varepsilon_{\gamma})\, \dot{\Theta}_i\, \dot{\eta}_{\Theta_i}\,\mathrm{d}t
= \Big[ I_i(\varepsilon_{\gamma})\dot{\Theta}_i \eta_{\Theta_i} \Big]_{t_1}^{t_2}
- \int_{t_1}^{t_2} I_i(\varepsilon_{\gamma})\, \ddot{\Theta}_i\, \eta_{\Theta_i}\,\mathrm{d}t 
\nonumber\\
\qquad\qquad
- \int_{t_1}^{t_2} I_i'(\varepsilon_{\gamma})\, \dot{\varepsilon}_{\gamma}\, \dot{\Theta}_i\, \eta_{\Theta_i}\,\mathrm{d}t,
\label{StationaryCondition2}
\end{gather}
\begin{gather}
\int_{t_1}^{t_2} M_{\alpha\beta}(\varepsilon_{\gamma})\, \dot{\varepsilon}_{\beta}\, \dot{\eta}_{\varepsilon_{\alpha}}\,\mathrm{d}t
= \Big[ M_{\alpha\beta}(\varepsilon_{\gamma})\dot{\varepsilon}_{\beta} \eta_{\varepsilon_{\alpha}} \Big]_{t_1}^{t_2}
- \int_{t_1}^{t_2} M_{\alpha\beta}(\varepsilon_{\gamma})\, \ddot{\varepsilon}_{\beta}\, \eta_{\varepsilon_{\alpha}}\,\mathrm{d}t 
\nonumber\\
\qquad\qquad
- \int_{t_1}^{t_2} M_{\alpha\beta}'(\varepsilon_{\gamma})\, \dot{\varepsilon}_{\gamma}\, \dot{\varepsilon}_{\beta}\, \eta_{\varepsilon_{\alpha}}\,\mathrm{d}t,
\label{StationaryCondition3}
\end{gather}

Based on the argumentation provided in Section~\ref{sectionTheoreticalPreliminaries}, we enforce vanishing variations at $t=t_1$ and $t=t_2$, i.e.
\begin{gather}\label{VanishingVariations}
\eta_{X_i}(t_1)=0,\quad \eta_{X_i}(t_2)=0,\nonumber\\[4pt]
\eta_{\Theta_i}(t_1)=0,\quad \eta_{\Theta_i}(t_2)=0, \nonumber\\[4pt]
\eta_{\varepsilon_{\alpha}}(t_1)=0,\quad \eta_{\varepsilon_{\alpha}}(t_2)=0,
\end{gather}
leading to vanishing square bracket terms. Finally inserting these results into Eq.~\ref{StationaryConditionVersionCurrentAppendix}, we obtain 
\begin{gather}
\delta S =
\int_{t_1}^{t_2} \Big(
- M_i \ddot{X}_i\, \eta_{X_i}
+ \tfrac{1}{2} I_i'(\varepsilon_{\gamma})\, \eta_{\varepsilon_{\gamma}}\, \dot{\Theta}_i^{\,2}
- I_i(\varepsilon_{\gamma})\, \ddot{\Theta}_i\, \eta_{\Theta_i}
- I_i'(\varepsilon_{\gamma})\, \dot{\varepsilon}_{\gamma}\, \dot{\Theta}_i\, \eta_{\Theta_i}
 \nonumber\\
\qquad\qquad
+\tfrac{1}{2} M_{\alpha\beta}'(\varepsilon_{\gamma})\, \eta_{\varepsilon_{\gamma}}\, \dot{\varepsilon}_{\alpha} \dot{\varepsilon}_{\beta}
- M_{\alpha\beta}(\varepsilon_{\gamma})\, \ddot{\varepsilon}_{\beta}\, \eta_{\varepsilon_{\alpha}}
- M_{\alpha\beta}'(\varepsilon_{\gamma})\, \dot{\varepsilon}_{\gamma} \dot{\varepsilon}_{\alpha}\, \eta_{\varepsilon_{\alpha}} \nonumber\\
\qquad\qquad
- U_\varepsilon'(\varepsilon_{\alpha})\, \eta_{\varepsilon_{\alpha}}
\Big)\,\mathrm{d}t.
\end{gather}

\section{Numerical parameters used in the validation examples}
\label{Appendix:ValidationParameters}

This appendix summarizes the numerical parameters used in the validation examples presented in Sections~\ref{analyticalModesSection} and \ref{NumericallyDerivedModesSection}.

\begin{table}[H]
\centering
\footnotesize
\caption{Input parameters for the single-particle single bending mode validation example. All quantities are expressed in voxel-based, dimensionless units.}
\label{table:InputsForBendingExample}

\setlength{\tabcolsep}{4pt}
\renewcommand{\arraystretch}{1.05}

\begin{tabular}{@{}
p{4.0cm}
>{\raggedright\arraybackslash\hyphenpenalty=10000\exhyphenpenalty=10000}p{4.0cm}
r
l
@{}}
\toprule
Category & Parameter & Value & Unit \\
\midrule
Geometry
& Particle length & 160 & voxels \\
& Particle thickness & 4 & voxels \\
& Particle width & 4 & voxels \\
& Level set grid spacing & 1 & voxel \\
\addlinespace

Modal
& Modal stiffness & $1\times10^{4}$ & -- \\
& Modal mass & $1\times10^{4}$ & -- \\
& Modal damping & $2\times10^{4}$ & -- \\
\addlinespace

Loading
& Maximum imposed displacement & 5 & voxels \\
& Loading rate & $5\times10^{-7}$ & voxel/step \\
\addlinespace

Integration
& Time step & $2.55012\times10^{-5}$ & -- \\
& Total steps & $1\times10^{7}$ & steps \\
\bottomrule
\end{tabular}
\end{table}

\begin{table}[H]
\centering
\footnotesize
\caption{Input parameters for the single-particle single numerical deformation mode validation example. The modal stiffness is state-dependent and calibrated from high-fidelity Abaqus simulations. All quantities are expressed in voxel-based, dimensionless units.}
\label{table:InputsForNumericalMode}

\setlength{\tabcolsep}{4pt}
\renewcommand{\arraystretch}{1.05}

\begin{tabular}{@{}
p{2.4cm}
>{\raggedright\arraybackslash\hyphenpenalty=10000\exhyphenpenalty=10000}p{4.2cm}
r
l
@{}}
\toprule
Category & Parameter & Value & Unit \\
\midrule
Geometry
& Particle radius & 10 & voxels \\
& Level set grid spacing & 1 & voxel \\
\addlinespace

Modal
& Modal stiffness & Calibrated from Abaqus & -- \\
& Modal mass & $1.0\times10^{5}$ & -- \\
& Modal damping & Proportional to stiffness & -- \\
\addlinespace

Loading
& Maximum imposed displacement & 1.6 & voxels \\
& Loading rate & $1.6\times10^{-7}$ & voxel/step \\
\addlinespace

Integration
& Time step & $6.47\times10^{-4}$ & -- \\
& Total steps & $1.0\times10^{7}$ & steps \\
\bottomrule
\end{tabular}
\end{table}

The nonlinear force--displacement relation defining the modal stiffness was obtained by extracting the indenter reaction force from the corresponding high-fidelity Abaqus simulation as a function of the generalized modal coordinate $\varepsilon$. The lookup table used in the deformable DEM simulation was defined by the discrete pairs
\[
{
\begin{aligned}
\varepsilon = \{&
0,\; 0.004,\; 0.008,\; 0.014,\; 0.023,\; 0.0365,\; 0.04156,\; 0.04916,\; 0.06055,\\
&0.07763,\; 0.08404,\; 0.09365,\; 0.10807,\; 0.11347,\; 0.12158,\; 0.13375,\\
&0.15199,\; 0.16
\},
\\
F_\varepsilon = \{&
0,\; 4.12,\; 11.95,\; 28.63,\; 63.39,\; 134.77,\; 167.32,\; 222.16,\; 317.71,\\
&493.45,\; 569.48,\; 694.49,\; 908.45,\; 996.90,\; 1138.99,\; 1373.51\\
&1777.76,\; 1974.62
\}.
\end{aligned}
}
\]

\section{Derivation of the semi-analytical wall force expression}\label{App:WallForceExpression}

The total normal force exerted by the six surrounding rigid walls on the particle pack is computed and tracked as a function of the imposed wall displacement. Owing to the frictionless and conservative nature of the system, the mechanical work performed by the external walls during compression must equal the total elastic energy stored within the deformable particles. Furthermore, under conditions of stress equilibrium and kinematic uniformity of the deformation modes within the compacted assembly, the global force--displacement response of the 27-sphere system can be inferred from the response of a single deformable sphere subjected to an equivalent compaction loading, as obtained from the corresponding FEM simulation.

Denoting by $u_w$ the imposed wall displacement and by $F_w(u_w)$ the total compressive force acting on the six walls, the external mechanical work reads
\begin{equation}
W_{\mathrm{ext}}=\int F_w(u_w)\,\mathrm{d}u_w,
\end{equation}
while the total elastic energy stored in the particle assembly is given by
\begin{equation}
W_{\mathrm{int}}=N\,U_\varepsilon(\varepsilon),
\end{equation}
where $N$ denotes the number of deformable spheres and $U_\varepsilon(\varepsilon)$ is the elastic potential energy of a single sphere under compaction, expressed in terms of the corresponding generalized modal amplitude.

Stress equilibrium, kinematic compatibility, and the symmetry of the cubic arrangement under uniform compaction yield
\begin{equation}
u_w = 3\,\varepsilon .
\end{equation}

Equating the work done by the walls and the elastic energy stored in the system, $W_{\mathrm{ext}} = W_{\mathrm{int}}$, gives
\begin{equation}
\int F_w(u_w)\,\mathrm{d}u_w = N\,U_\varepsilon\!\left(\tfrac{u_w}{3}\right).
\end{equation}
Differentiating both sides with respect to $u_w$ yields
\begin{equation}
F_w(u_w) = \frac{N}{3}\,U_\varepsilon'\!\left(\tfrac{u_w}{3}\right),
\end{equation}
where $U_\varepsilon'(\varepsilon)=\mathrm{d}U_\varepsilon/\mathrm{d}\varepsilon$ denotes the generalized internal elastic force, denoted by $F_{\mathrm{int}}(\varepsilon)$, associated with the single-sphere compaction problem and conjugate to the generalized modal amplitude $\varepsilon$, evaluated at a rescaled argument corresponding to the imposed wall displacement. For the representative cluster considered here ($N=27$), this expression reduces to
\begin{equation}
\boxed{
F_w(u_w) = 9\,F_{\mathrm{int}}\!\left(\tfrac{u_w}{3}\right)
}.
\end{equation}

\bibliographystyle{elsarticle-num}   
\bibliography{ref}                   

\end{document}